\documentclass[twocolumn]{aastex631}

\usepackage{soul}
\usepackage{slashbox}

\shorttitle{Photochemistry of gaseous exoplanets}

\graphicspath{{./}{figures/}}

\begin{document}

\title{EXTREME ULTRAVIOLET AND X-RAY DRIVEN PHOTOCHEMISTRY \\ OF GASEOUS EXOPLANETS}

\author[0000-0002-9824-2336]{Daniele Locci}
\affiliation{INAF - Osservatorio Astronomico di Palermo, P.za Parlamento 1, 90134 Palermo, Italy}

\author[0000-0002-9882-1020]{Antonino Petralia}
\affiliation{INAF - Osservatorio Astronomico di Palermo, P.za Parlamento 1, 90134 Palermo, Italy}

\author[0000-0002-9900-4751]{Giuseppina Micela}
\affiliation{INAF - Osservatorio Astronomico di Palermo, P.za Parlamento 1, 90134 Palermo, Italy}

\author[0000-0001-5154-6108]{Antonio Maggio}
\affiliation{INAF - Osservatorio Astronomico di Palermo, P.za Parlamento 1, 90134 Palermo, Italy}

\author[0000-0002-3127-8078]{Angela Ciaravella}
\affiliation{INAF - Osservatorio Astronomico di Palermo, P.za Parlamento 1, 90134 Palermo, Italy}

\author[0000-0001-7480-0324]{Cesare Cecchi-Pestellini}
\affiliation{INAF - Osservatorio Astronomico di Palermo, P.za Parlamento 1, 90134 Palermo, Italy}

\begin{abstract}
The interaction of exoplanets with their host stars causes a vast diversity in bulk and atmospheric compositions, and physical and chemical conditions. Stellar radiation, especially at the shorter wavelengths, drives the chemistry in the upper atmospheric layers of close orbiting gaseous giants, providing drastic departures from equilibrium. In this study, we aim at unfolding the effects caused by photons in different spectral bands on the atmospheric chemistry, with particular emphasis on the molecular synthesis induced by X-rays. This task is particularly difficult because the characteristics of chemical evolution emerge from many feedbacks on a wide range of time scales, and because of the existing correlations among different portions of the stellar spectrum.  

The weak X-ray photoabsorption cross-sections of the atmospheric constituents boost the gas ionization to pressures inaccessible to vacuum and extreme ultraviolet photons. Although X-rays interact preferentially with metals, they produce a secondary electron cascade able to ionize efficiently hydrogen and helium bearing species, giving rise to a distinctive chemistry.
\end{abstract}


\submitjournal{PSJ}

\section{Introduction}
Exoplanets form and evolve under the influence of their host stars. In the process, planetary atmospheres naturally arise and modify under selective environmental constraints, providing an astonishing diversity in compositions, and physical and chemical conditions. Our knowledge of exoplanets' atmospheres has improved dramatically over the last two decades (e.g., \citealt{Tsiaras19,Giacobbe21}), spanning a broad range of planetary types, comprising also gas and ice giants, and super-Earths. Just as it occurred for solar system planets, an esoteric field of research is now becoming a major area of interest to physicists and chemists.

Theoretical models are beginning to yield important insights into the chemistry of exoplanetary atmospheres (e.g., \citealt{Venot15}). While the chemical composition is at the equilibrium in deep atmospheric layers \citep{Madhusudhan16}, kinetic processes drive drastic departures from equilibrium in the upper regions of an atmosphere, and may also involve the possibility of escape of its constituents to space (e.g., \citealt{Koskinen14,King18}). The main kinetic mechanisms affecting equilibrium chemistry are transport-induced quenching and photochemistry. Here, we use the term photochemistry in a broad sense, including the effects of ionizing radiation (usually called radiation chemistry).  Photochemical reactions dominate in the upper atmospheric layers, in a range of pressure determined by the composition, and the stellar illumination. 

In this work, we are mainly interested in the effects of the stellar high energy radiation on the upper atmospheric layers of gaseous giants, occurring for pressures lower than $P \sim 10^{-2} - 10^{-3}$~bar. Some previous studies investigate the impact of molecular photodissociation on chemical abundances (e.g., \citealt{Moses11,Molaverdikhani19}), while others include the effects of photoionization processes and ion-neutral chemistry (e.g., \citealt{Garcia07,Erkaev13,Bourgalais20}), although in such a context, just a few contain extended chemical networks (e.g., \citealt{Barth21}). Not many works include detailed descriptions of the secondary electron cascade (e.g., \citealt{CCP06,Shematovich14}). Additional ionizing sources, such as cosmic rays and stellar energetic particles are addressed in  \citet{Airapetian16,Airapetian17} and \citet{Barth21}, while lightning and charge processes in \citet{Helling19}.

A key element in the accurate description of photochemistry is the representation of the illuminating radiation field and its energy dependence, including the energy tail extended into the X-ray domain. Stellar radiation may present correlations among the intensities in various spectral ranges (e.g., \citealt{Sanz-Forcada11,King18}), and significant variations with the stellar age impacting differently at different energies \citep{Micela02,Ribas05}. Thus, it is not surprising the wide dispersion in radiation fields adopted in the literature, either in spectral ranges and shapes. Some authors assume observed spectra of specific stars e.g., using {\it HST} and {\it XMM-Newton/Swift} telescope \citep{Barth21}, or the {\it PHOENIX} library (e.g., \citealt{Kitzmann18}); others scale the solar spectrum by means of coronal models (e.g., \citealt{Sanz-Forcada11}), as done by \citet{Chadney15}, or exploit spectra taken from the {\it Virtual Planetary Laboratory}, to investigate the effect of an increased stellar activity \citep{Shulyak20}. Stellar X-ray emission may also be simulated through thermal bremsstrahlung \citep{Lorenzani01} and thermal emission of hot plasmas (e.g., \citealt{CCP09,Locci18}).

Because of their high energies, Extreme UltraViolet (EUV) and X-ray photons produce phenomena that cannot be caused in any other of the lower energy bands, regardless of their larger fluxes. In atmospheres with solar-like composition, the main interactions of EUV photons occur in the very upper layers, while X-rays owing to their smaller absorption cross-sections, may penetrate much further downward. A unique feature of ionizing radiation is that all the relevant processes are dominated by a secondary, low-energy electron cascade generated by the primary photo-electron \citep{Maloney96,Arumainayagam21}. The effects produced by secondary electrons are by far more important than the corresponding ionization, excitation, and dissociation events caused directly by X-rays (e.g., \citealt{Locci18}). This is a consequence of the large primary photoelectron energies. Such secondary cascade keeps ionizing the gas \citep{CCP06,Johnstone18}. At the end of the energy degradation process, the residual energy unable of providing further excitation goes into the gas heating \citep{Dalgarno99,CCP06}. The global chemical effect is a substantial rise in the ionization level extending deeply into the atmosphere. When the electron fraction in the gas exceeds few percents, electrons loose preferentially their energies through electron-electron Coulomb interactions \citep{Dalgarno99}, so that too large radiation fluxes produce rather weak non-thermal effects. 
 
In Section~\ref{r&c} we describe the model and the data needed to describe specific representations. In Section \ref{res} we present the results for a model defined by standard assumptions in the main physical and chemical parameters (e.g., X-ray luminosity and metallicity), and we compare them to those stemming from variations in such fiducial values. In the last Section we discuss the results and outline our conclusions.

\section{Chemistry and Radiation} \label{r&c}
We have developed a one-dimensional thermochemical and photochemical kinetics model to describe the vertical chemical profiles of 128 selected species, including electrons (see Table~\ref{tone}). All the neutral species listed in the Table possess singly charged, positive counterparts. No anions are included in the network. The selected species consist of 5 elements ~\textemdash H, He, C, N, and O\textemdash~ coupled through a network of 1978 chemical reactions. The reaction inventory contains bimolecular, termolecular, thermodissociative, ion-neutral, and photochemical reactions. These latter include photodissociations, mainly due to UV and EUV radiation, and photoionizations by EUV radiation and X-rays. Specifically we consider 617 neutral-neutral reactions and 912 neutral-ion reactions for a total of 1529 bimolecular reactions, 40 termolecular reactions, 70 thermodissociative reactions, 127 photochemical reactions including either dissociation and ionization processes; 212 neutral-neutral and termolecular reactions have been reversed. The list of reactions is reported in the Supplementary Materials available at \dataset[10.5281/zenodo.5638699]{https://doi.org/10.5281/zenodo.5638699}.

\begin{table*}[t]
\caption{Chemical species}
\begin{tabular}{lc} 
\hline
Atoms & Species \\
\hline 
H, He &  H, He, H$_2$  \\
H, O & O, O(1D), O$_2$, O$_3$, OH, H$_2$O, HO$_2$, H$_2$O$_2$ \\
H, O, C & C, C$_2$, CH, CH$_2$, CH$_3$, CH$_4$, C$_2$H$_2$, C$_2$H, C$_2$H$_3$, C$_2$H$_4$, C$_2$H$_5$, C$_2$H$_6$, HCO, CO, CO$_2$, H$_2$CO, C$_2$HO,\\
&  CH$_3$O, CH$_3$O$_2$ ,CHO$_2$, CH$_2$O$_2$, CH$_4$O$_2$, CH$_2$OH, CH$_3$OH, C$_2$H$_3$O, C$_2$H$_2$O, C$_2$H$_4$O, C$_2$H$_5$O  \\
H, O, C, N & N, N$_2$, NH, NH$_2$, NH$_3$, N$_2$H$_3$, N$_2$H$_4$, NO, N$_2$O, NO$_2$, NO$_3$, N$_2$O$_5$, HNO, HNO$_2$,HNO$_3$, HCN,\\
&  CN, CNO, HCNO, HNCO, HCONH$_2$, CH$_5$N, C$_2$H$_2$N \\
additional ions & H$_3^+$, HeH$^+$, H$_3$O$^+$ \\
\hline
\end{tabular}
\label{tone}
\end{table*}

The chemical evolution is described by the system of differential equations
\begin{equation}
\frac{dn_i}{dt} = P_i-n_iL_i
\end{equation}
where $n_i$ is the number density of the $i-$th species, and $P$ and $L$ are production and destruction terms referring to all chemical and physical processes that produce and destroy the $i-$th species. They are therefore functions of all the species included in the network of chemical reactions. The time derivative is to be understood as comoving. Although, the density-based solver may be coupled to flow and energy equations, we do not consider any motion within the fluid, with the chemistry evolving in a static atmosphere. While vertical mixing (and other motions) may be indeed important in the chemical balance of planetary atmospheres (e.g., \citealt{Moses11,Agundez14}), we choose not to include atmospheric dynamics (and in general, any form of non-chemical perturbation), to highlight the role of dissociating and ionizing radiation as a source of chemical disequilibrium, and primarily the relative importance of different spectral energy bands. 

\subsection{Photochemistry} \label{subsec:photo}
Photochemical rates describing both dissociation and ionization are computed as follows
\begin{equation}
\beta(r) = \int_{E_{\rm th}}^\infty \sigma(E) F (E,r) dE
\end{equation}
where $\sigma(E)$ is the cross-section associated with a specific photoprocess, $E_{\rm th}$ the corresponding energy threshold, and $F(E,r)$ the radiative flux at the altitude $r$ in the atmosphere. Ionizations must be supplemented by an additional term including the effects of the secondary electron cascade
\begin{equation}
\beta_{\rm sec} = \int n_{\rm sec}(E) v(E) \sigma_e(E) dE
\label{sec}
\end{equation}
(e.g., \citealt{Adam11,Locci18}) where $v(E)$ is the electron velocity, $n_{\rm sec}(E)$ the absolute number distribution of secondary electrons, and $\sigma_e(E)$ the energy-dependent electron impact ionization cross-section. Typically, such cross-sections have an asymmetric bell-shape, with a threshold around 10 eV, a peak at $100 - 300$~eV, while declining to small values around 1 keV (e.g., \citealt{Hudson04}). Equation~(\ref{sec}) includes also the contribution of Auger electrons (see \citealt{Locci18}). The number distribution of secondary electrons depends on the inverse of the mean energy per ion pair $W$ (e.g., \citealt{CCP92}), which is the initial energy of the photo-electron divided by the number of secondary ionizations produced as the  particle comes to rest. In a gas of solar-like composition, the limiting value of $W$ at high energies is $\sim 35$~eV, while it is infinite at the ionization threshold \citep{Dalgarno99}. For this reason, the contribution of the secondary electron cascade decreases sharply when the energy approaches the Lyman continuum. We compute $W$ as a function of the energy of the primary photo-electron, for different abundance ratios and electronic concentrations as done in \citet{CCP06}, and store them in a look-up table. 

We use measured or calculated photoabsorption cross-sections (see e.g., the on-line database Photoionization/Dissociation Rates\footnote{https://phidrates.space.swri.edu}, \citealt{Huebner15}). In those cases data were not available, we approximate molecular X-ray absorption cross-sections by means of the cross-sections of the constituent atoms \citep{Maloney96,Yan97}. The individual atomic cross-sections are taken from the compilation of \citet{Verner96}. Known total  cross-sections provide support to such an assumption. Photodestruction cross-sections in lower-energy spectral ranges have been taken from existing compilations (e.g., KIDA, the Kinetic Database for Astrochemistry, \citealt{Wakelam15}).

\subsection{Radiative Transfer}
\begin{figure}
\centering
\includegraphics[width=9cm]{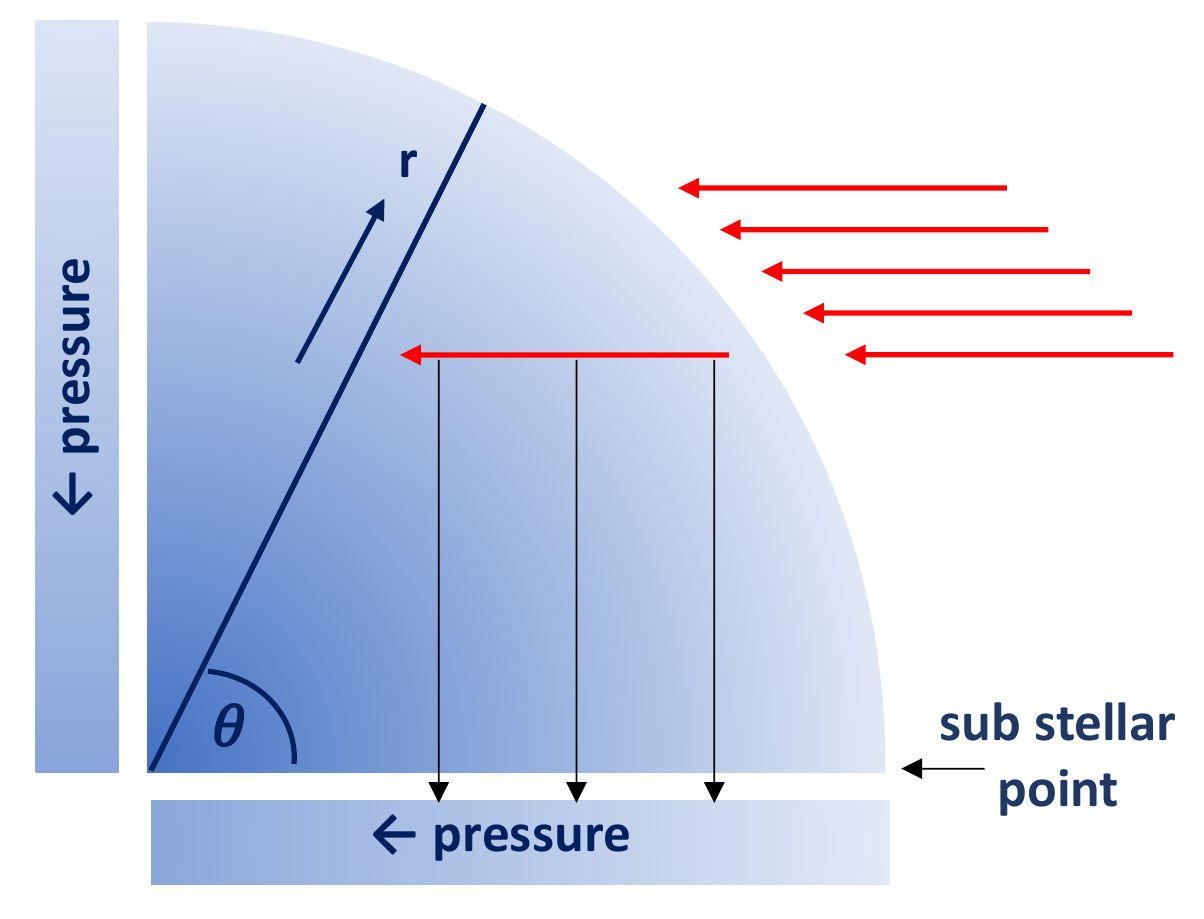}
\caption{The geometry of the radiation transfer in our model, with the incoming radiation travelling horizontally. The zenith angle $\theta$ identifies the radial path along which the chemistry is computed. The vertical black arrows indicates the pressures sampled along the photon path.}
\label{fone}
\end{figure}

We consider a one-dimensional geometry, in which a stratified, cloud-free atmosphere is illuminated by a stellar photon flux in the range between 3 and 10000~eV, and we derive the chemistry along a direction defined by the zenith angle $\theta$ (Figure~\ref{fone}).

The radiation at any altitude $r$ is obtained as 
\begin{equation}
F (E,r)=F_\star(E) e^{-\tau(E,r)}
\label{expo} 
\end{equation}  
where $F_\star(E)$ is the stellar flux impinging at the (arbitrary) outer boundary of the atmosphere, and $\tau (E,r)$ the total atmospheric optical depth at the altitude $r$
\begin{equation}
\tau (E,r)=  \sum_i \sigma_i(E) N_i(r)
\label{tauh}
\end{equation}
In equation (\ref{tauh}), $\sigma_i(E)$ and $N_i(r)$ are the total absorption cross-section, and the column density of the $i-$th species, respectively; the latter depends on the zenith angle, and it is obtained integrating along the horizontal direction identified by the corresponding radiation path, as illustrated in Figure~\ref{fone}. 

\subsection{Input values and boundary conditions}
We assume a planet with mass equal to half Jupiter mass, orbiting around a Sun-like star with solar bolometric luminosity. The adopted stellar spectrum is a mosaic from several sources. Low energy wavebands, up to the Lyman continuum are described by a {\it PHOENIX} library model of a G-type star \citep{Husser13}, plus a Lyman-$\alpha$ emission line, whose intensity is related to the X-ray luminosity \citep{Linsky20}. The spectral luminosity (ergs~s$^{-1}$~eV$^{-1}$) in the X-ray domain ($0.1-10$~keV), ${\cal L}_{\rm X}$, is modelled exploiting Raymond-Smith models for the thermal emission of hot plasmas \citep{Raymond77}. The total X-ray luminosity (in ergs~s$^{-1}$)
\begin{equation}
L_{\rm X} = \int_{0.1~\rm keV}^{\rm 10~keV}  {\cal L}_{\rm X} (E,T_{\rm X}) dE
\label{bright}
\end{equation}
and the hardness of the spectrum (i.e. its temperature, $T_{\rm X}$), are free parameters. 

EUV radiation ($13.6-100$~eV) is difficult to determine observationally, and its quantification is frequently based on semiempirical models extended into the EUV domain from either X-ray or UV observations (e.g., \citealt{Chadney15,Fontenla15}). This can be done reconstructing the EUV emission from Lyman-$\alpha$ measurements \citep{Linsky14}, or through solar data from the {\it TIMED/SEE} mission \citep{Chadney15}, extrapolating them to {\it XMM-Newton} and {\it Chandra} observations \citep{King18}. We follow the approach put forward by \citet{Sanz-Forcada11}, who derived an expression relating the EUV and X-ray luminosities, based on synthetic coronal models for a sample of main sequence stars. Since in the EUV energy range the spectral shape is rather uncertain, we adopt a flat spectral distribution. 

\begin{figure}
\centering
\includegraphics[width=9cm]{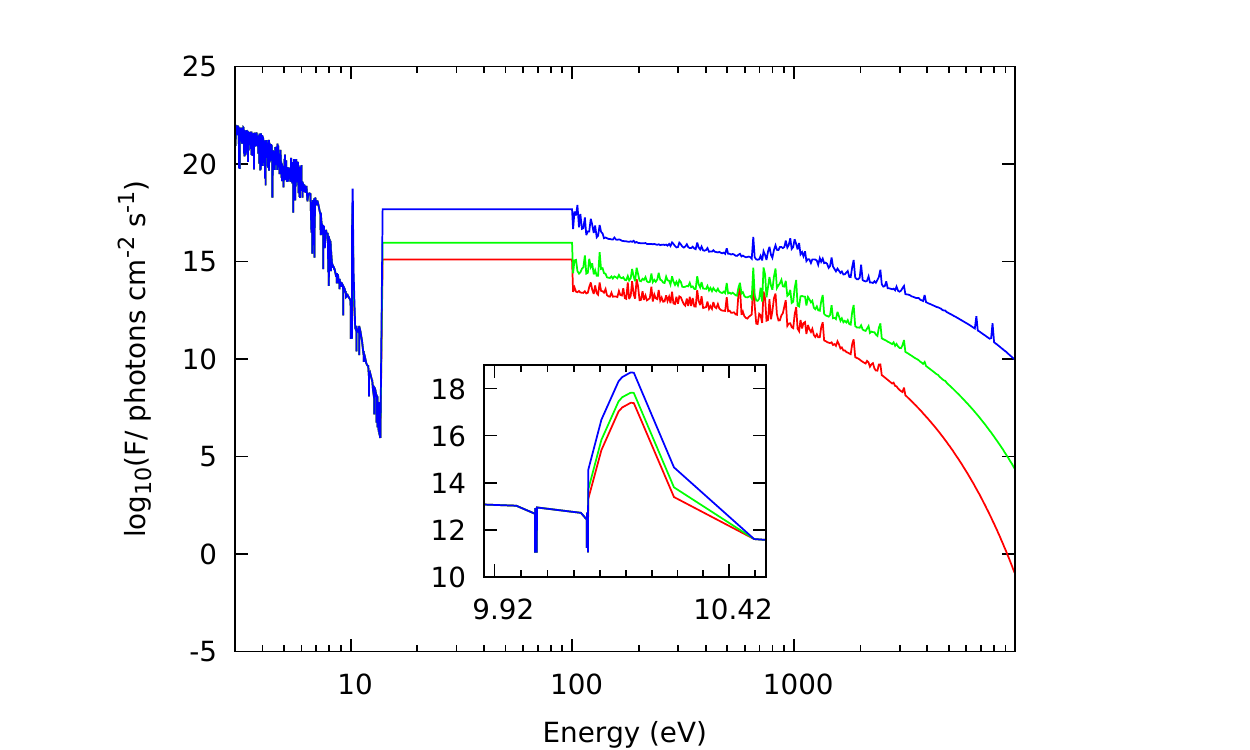}
\caption{Stellar emission $F$ in the energy range between 3 eV and 10 keV. Red line: $L_{\rm X} = 1 \times 10^{27}$~ergs~s$^{-1}$, $T_{\rm X} = 0.3$~keV; green line: $L_{\rm X} = 1 \times 10^{28}$~ergs~s$^{-1}$, $T_{\rm X} = 0.5$~keV; blue line: $L_{\rm X} = 1 \times 10^{30}$~ergs~s$^{-1}$, $T_{\rm X} = 1$~keV. In the EUV band, we assume a flat spectral shape related to the X-ray luminosity by the \citet{Sanz-Forcada11} relation. In the inset, we report the portion encompassing the Lyman$-\alpha$ line, that is also related to $L_{\rm X}$ \citep{Linsky20}; the asymmetry in the line profile stems out of the resolution with which we sample the spectrum.}
\label{ftwo}
\end{figure}

Ultimately, we adopt: (1) a hot plasma thermal emission for photon energies comprised between 1~keV and 100 eV (X-rays), with the integrated luminosity, equation (\ref{bright}), and the plasma temperature, $T_{\rm X}$ being free parameters; (2) a constant spectrum (not dependent on photon energy) in the EUV band $100 < E < 13.6$~eV, whose luminosity scales directly with the X-ray luminosity according to the \citet{Sanz-Forcada11} relation, $L_{\rm EUV} = 6.31 \times 10^4 \, L_{\rm X}^{0.86}$; (3) a Lyman$-\alpha$ profile at 10.2~eV, whose intensity is related to the X-ray luminosity by the expression $L_{\rm Ly \alpha} = 4.57 \times 10^{16} \, L_{\rm X}^{0.43}$ derived from data reported in \citet{Linsky20}; (4) a {\it PHOENIX} G-type star model providing the flux in the UV spectral range between 3 and 13.6~eV. This last portion of the stellar emission is the only one depending on the star spectral type. In Figure \ref{ftwo}, we report the adopted illuminating radiation field, under some assumptions  on  the  brightness  and  hardness  of the X-ray component.

The temperature profile is not calculated self-consistently within the code. We assume an isothermal atmosphere, where the chosen value is a free parameter (e.g., the planetary equilibrium temperature). While X-rays may give an important contribution to the heating of hydrogen-rich planetary atmospheres (e.g., \citealt{CCP09}), we make such an assumption, to avoid that temperature variations can influence, and entangle the impact of different spectral bands on the chemical profiles. We consider a range in pressure extending from $P = 1 \times 10^3$ to $1 \times 10^{-11}$~bar. For each pair of values $P-T$, remaining derived atmospheric variables, e.g., density and altitude, are obtained imposing hydrostatic equilibrium.

\begin{figure}
\centering
\includegraphics[width=9cm]{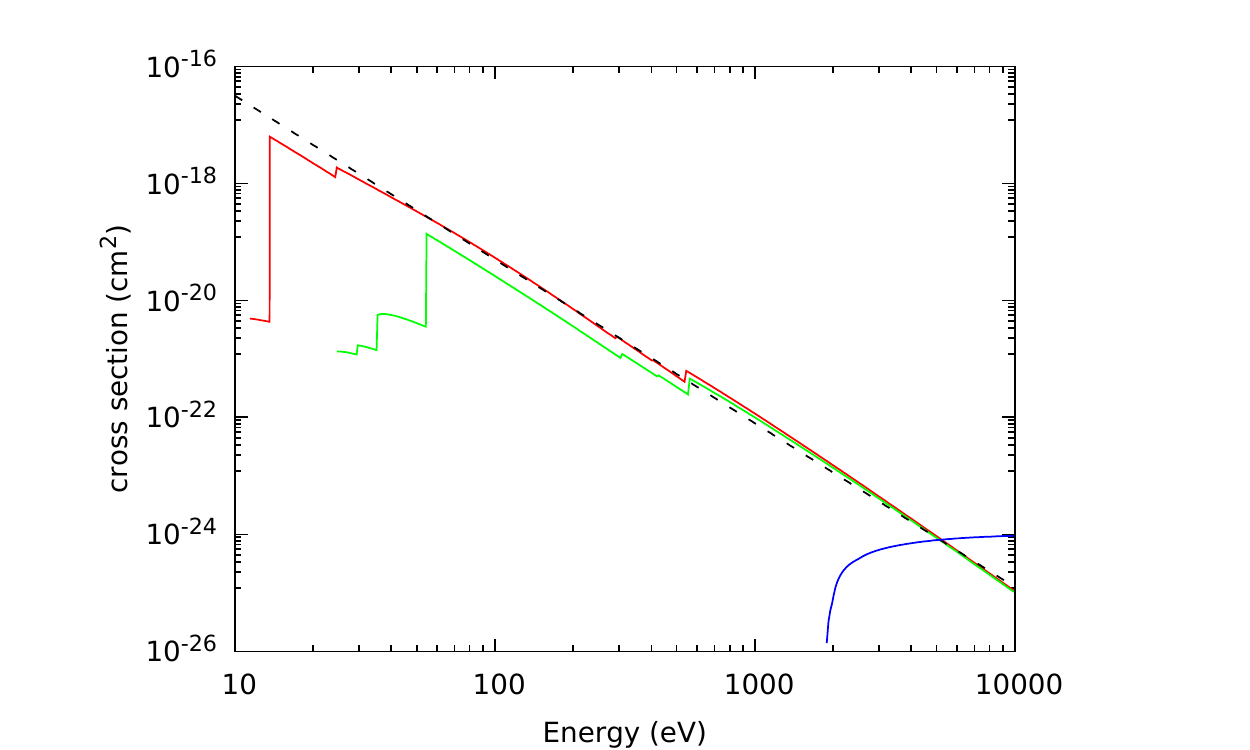}
\caption{Total photoionization cross-section (cm$^2$) as a function of the energy (eV) of the incoming photon. Red solid line: neutrals; green solid line: cations; blue solid line: Compton ionization cross-section \citep{Locci18}; black dashed line: straight line approximation, $\sim E^{-2.8}$.}
\label{fthree}
\end{figure}

We assume a solar chemical composition, with concentrations taken from the compilation of \citet{Asplund09}. As initial conditions, we take the gas in each layer to be neutral and in atomic form. In Figure \ref{fthree}, we report the total photoionization cross-section, computed considering all the elements present in the atmospheric gas either in neutral and singly ionized atomic forms. This quantity is close to the photoabsorption cross-section in the upper end of the EUV energy range and in the X-ray domain. Such cross-section scales with the energy as $\sigma (E) \propto E^{-\gamma}$ \citep{Maloney96}, with $\gamma = 2.8$. As a consequence, photons with larger energies penetrate deeper into the atmosphere.

The zenith angle is chosen to be $\theta = 60^\circ$, considered a good approximation for the globally averaged profile \citep{Johnstone18}. We set the total X-ray luminosity to $L_{\rm X} = 1 \times 10^{28}$~ergs~s$^{-1}$, and we select the spectral shape corresponding to a plasma temperature $T_{\rm X} = 0.5$~keV (see \citealt{Locci18}). For the atmospheric temperature we use the fiducial value $T = 1000$~K, consistent with a planetary orbital distance $d_{\rm P} = 0.045$~au. Finally, we do not include mass flux entering or leaving the system at the top and bottom boundaries. 

All the assumptions reported above identify our reference model (hereafter RF model). In the following, we will present the molecular vertical distribution for the RF model. Next, we will compare them to those obtained varying some critical stellar and atmospheric parameters (see Table~\ref{ttwo}). Of particular interest are those quantities affecting directly the transfer of high energy photons within the atmosphere, i.e. the stellar activity, and the metallicity. In order to understand the impact of different radiation energy ranges, we include some pathological unrealistic cases e.g., suppressing the EUV spectral band or the effects of the secondary electron cascade.
\begin{table*}
\caption{Models and model parameters}
\begin{tabular}{lccccl} 
\hline
model & $L_{\rm X}$ & $T_{\rm X}$ & EUV$^\dag$ & $Z/Z_\odot$ & notes \\ 
&($\times 10^{28}$~ergs~s$^{-1})$ & (keV) & (Y/N) & &   \\ \hline
& & & &  & \\
RF & 1 & 0.5 & Y & 1 &   reference (default) \\
LA & 0.1 & 0.3 &  &  &  low stellar activity \\
HA & 100 & 1.0 &  &  &  high stellar activity \\
NX & 0   &  & Y     &    & no X-rays \\
NE &    &  & N  &       & no EUV radiation \\
UV & 0   &  & N &       & just near and vacuum UV \\
NS &    &  &   &       & no secondary electrons \\
LM &    &  &   &      0.1  & low metallicity \\
HM &    &  &   &      10  & high metallicity \\
& & &  & & \\
\hline 
\end{tabular}
\flushleft
$\dag$ EUV and X-ray luminosities are related through the empirical scaling law given in \citet{Sanz-Forcada11}; thus, a high X-ray luminosity corresponds to a high stellar activity in both the EUV and X-rays spectral bands.
\label{ttwo}
\end{table*}

\section{Results} \label{res}
\begin{figure}
\centering
\includegraphics[width=10cm]{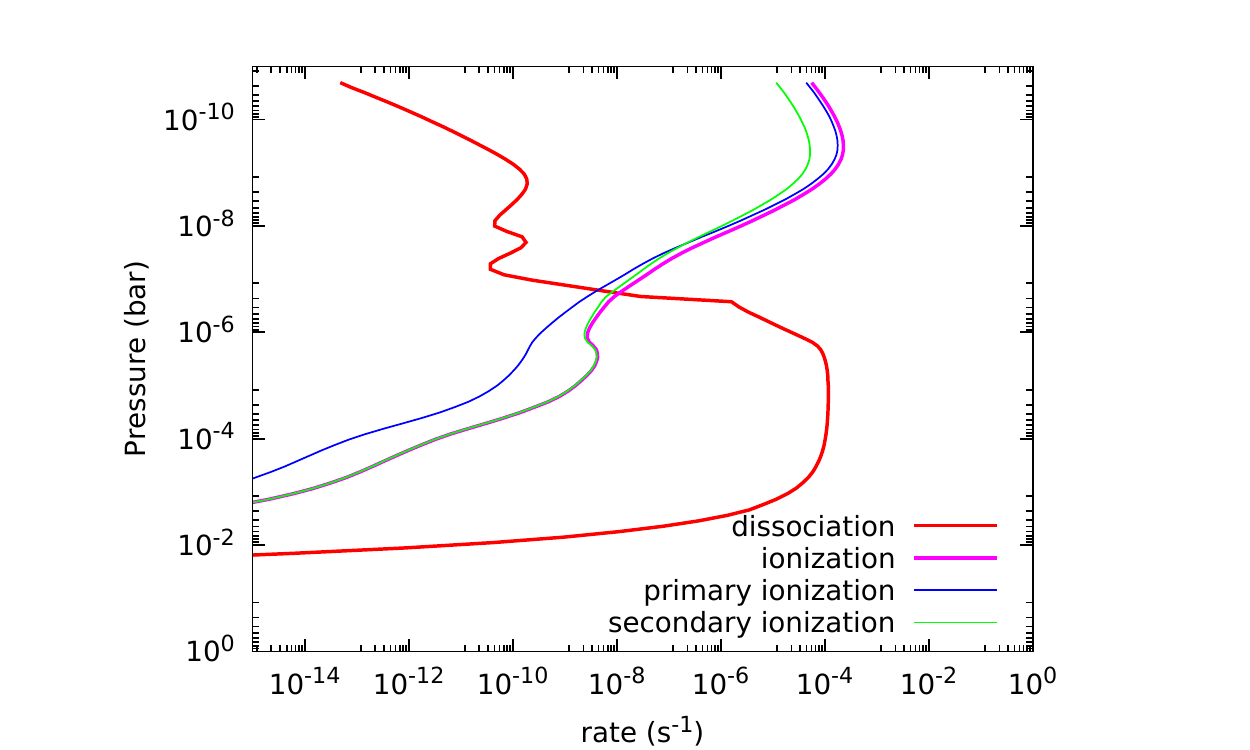} \\ 
\includegraphics[width=10cm]{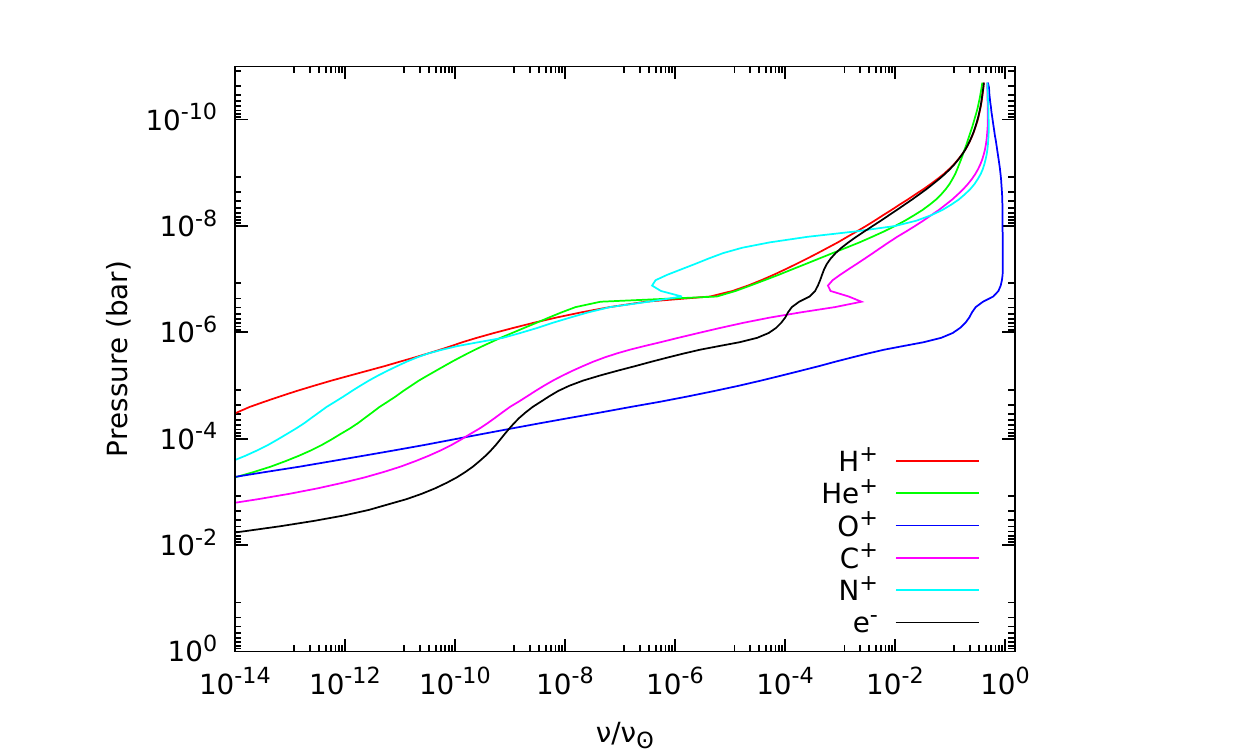}
\caption{Photochemical rates, and ion and electron profiles in the RF model. Upper panel: trends of different radiation-induced exit rates; lower panel: vertical distribution of normalized volume mixing ratios of atomic ions and electrons.}
\label{ffour}
\end{figure}

The ionization rate and the electron production, together with the rate of molecular dissociation (or dissociative ionization) are direct manifestations of the impact of the stellar energetic radiation on the atmospheric gas. We will describe the results proceeding from the top to the bottom of the atmosphere, i.e. for decreasing importance of photochemical effects. Figure~\ref{ffour}a shows that in the uppermost layers, gas ionization is dominated by radiation in the UV range, through valence ionization of atomic carbon (11.3~eV), and in the EUV spectral band, mainly via H (13.6~eV) and He (24.6~eV) ionizations. In the upper atmospheric layers the ionization is largely primary, while deeper down in the atmosphere, $P \ga 1 \times 10^{-8}$~bar, secondary ionization dominates, tracing the outbreak of X-rays. Although secondary ionization rates are lower than those produced by primary ionization, such residual ionization has a strong impact on molecular chemistry, that otherwise would be mostly driven by neutral-neutral reactions. This is a direct consequence of the much deeper penetration of X-ray radiation in atmospheres with solar-like composition. Photochemical effects decline significantly at pressures larger than $\sim 10^{-3}-10^{-2}$~bar. The vertical plateau in the cumulative dissociation rate marks the appearance of molecular species, that follows the drop of ionization. 

In Figure~\ref{ffour}b we show the vertical distribution of volume mixing ratios $\nu$ of atomic ions, normalized to the total concentrations of the corresponding elements, $\nu_\odot$. In a totally ionized gas, all normalized ratios tend to $\nu/\nu_\odot = 0.5$, a value that include the electron contribution. In Figure~\ref{ffour}b, it is also displayed the electron concentration, that follows closely the proton profile at the lowest pressures. Increasing the pressure, going deeper into the atmosphere, oxygen is the major repository of the ionization, that is mainly induced by EUV radiation up to pressures $P \sim 10^{-7}$ when X-rays take over, ejecting core (and Auger) electrons from the $K-$shells of C (0.28 keV), N (0.40 keV), and O (0.53 keV), spotted by the spikes (or bumps) in the C, N, and O profiles at pressures around $P \sim 10^{-6}$~bar. This is confirmed by the onset of secondary ionization occurring approximately at the same altitudes (as shown in Figure~\ref{ffour}a). Descending further, oxygen keeps controlling the ionization through molecular ions (see next Section), up to pressures $P \sim 10^{-3} - 10^{-2}$~bar, at which cloud formation should occur \citep{Madhusudhan19}. Such trends reflect the behaviour of the ionization cross-sections, with oxygen possessing the largest ones in both the EUV and X-ray bands.
\begin{table*}
\caption{Ion/neutral pressure crossing points (bar)}
\begin{tabular}{l|cccccccc} 
\backslashbox{atom}{model} &\makebox[3em]{RF} &\makebox[3em]{LA} &\makebox[3em]{HA} &\makebox[3em]{LM} &\makebox[3em]{HM} &\makebox[3em]{NX} &\makebox[3em]{NE$^\dag$} &\makebox[3em]{NS}  \\ \hline 
& & & & & & & & \\
H & 1.3(-10)$^\ddag$ & 2.0(-11) & 1.8(-9) & 1.6(-10) & 1.3(-10) & 1.3(-10) & & 1.3(-10) \\
He & 8.9(-11)\phantom{$^\dag$} & 2.0(-11) & 3.1(-9) & 1.6(-10) & 8.9(-11) & 8.9(-11) & & 8.9(-11) \\
C & 6.9(-10)\phantom{$^\dag$} & 1.4(-10) & 5.4(-9) & 5.5(-10) & 7.0(-10) & 6.9(-10) & & 6.9(-10) \\
N & 8.5(-10)\phantom{$^\dag$} & 2.1(-10) & 6.7(-9) & 8.4(-10) & 8.6(-10) & 8.5(-10) & & 8.5(-10) \\
O & 8.3(-7)\phantom{$^\dag$} & 8.3(-8) & 1.3(-4) & 1.7(-4) & 3.4(-7) & 6.5(-8) & 2.0(-5) & 2.0(-5) \\
& & & & & & & & \\
\hline
\end{tabular}
\flushleft
$\dag$ a blank space indicates that the ion concentration is lower than that of the corresponding neutral throughout the atmosphere; $\ddag$ 1.3(-10) = $1.3 \times 10^{-10}$.
\label{tthree}
\end{table*}

\begin{figure}
\centering
\includegraphics[width=10cm]{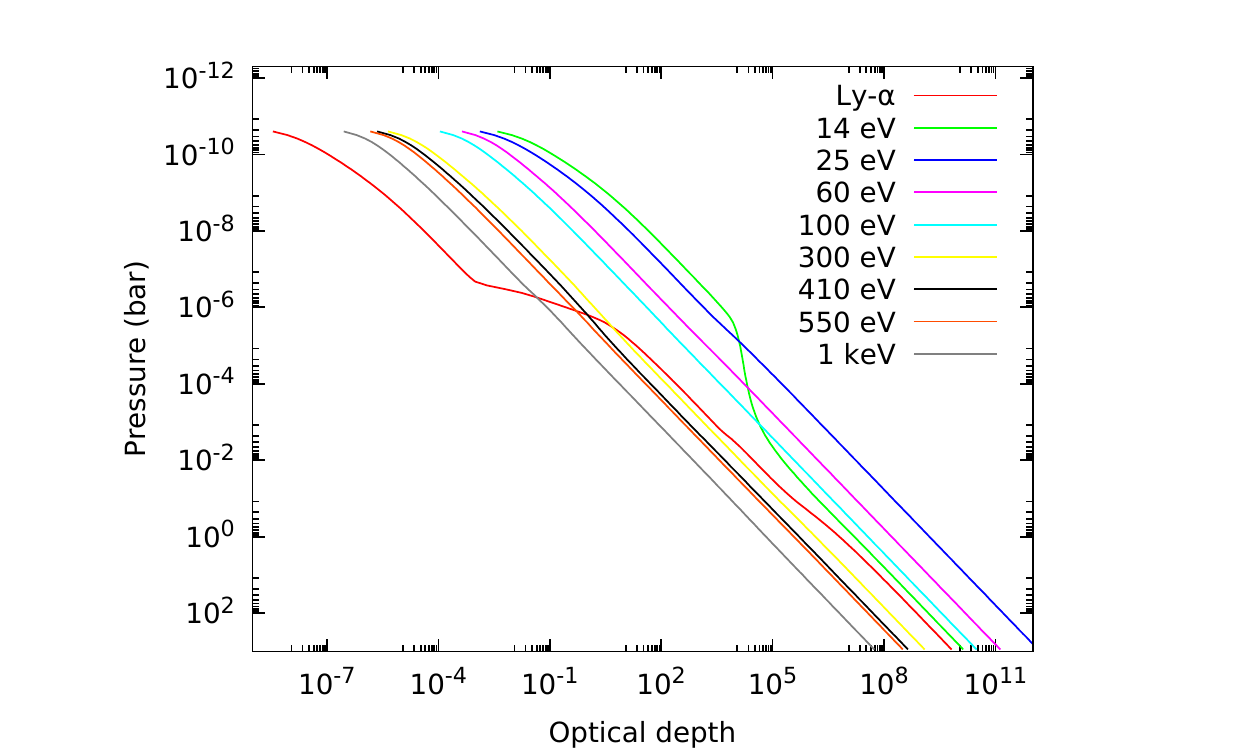} \\
\includegraphics[width=10cm]{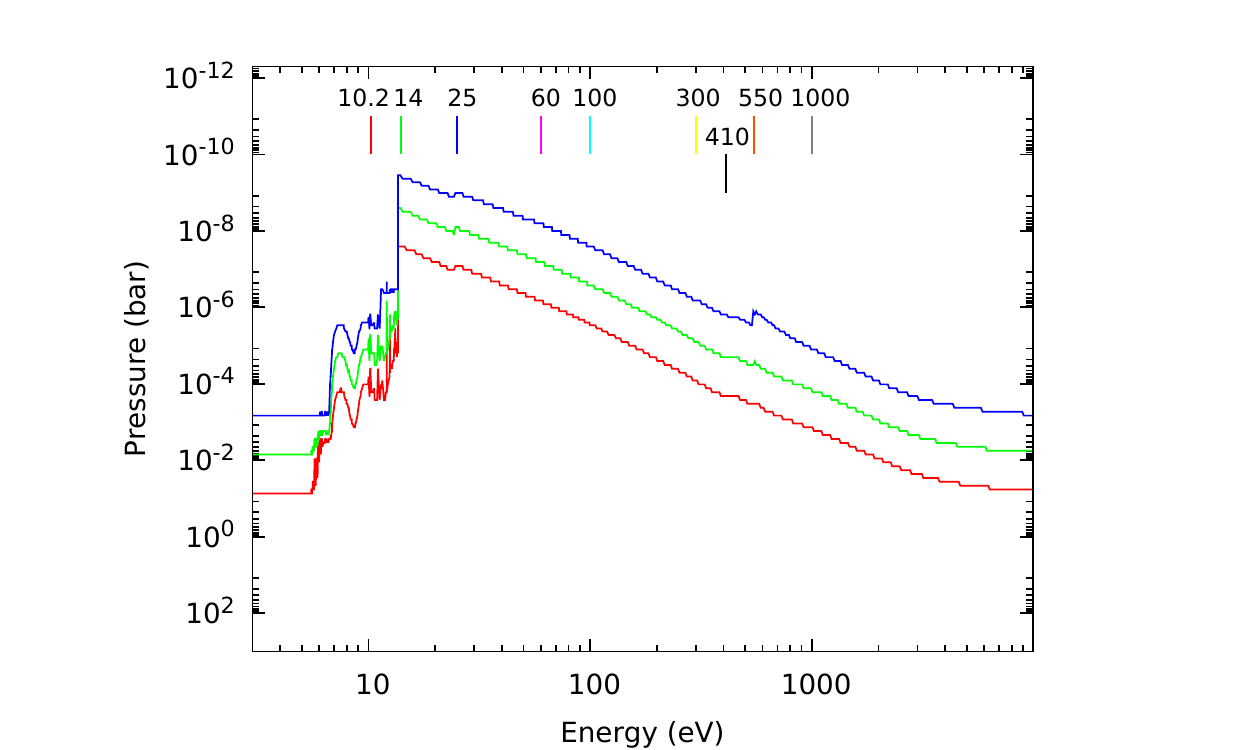}
\caption{Optical depths of photons of different energies. Top panel: distribution along the vertical pressure profile; Bottom panel: pressure level at which $\tau (E) = 1$ (blue line), 10 (green line), and 100 (red line). Labels indicate the set of photon energies (eV) exploited in the top panel.}  
\label{ffive}
\end{figure}

\begin{figure*}
\begin{tabular}{cc} 
\hspace{-1cm}
\includegraphics[width=10cm]{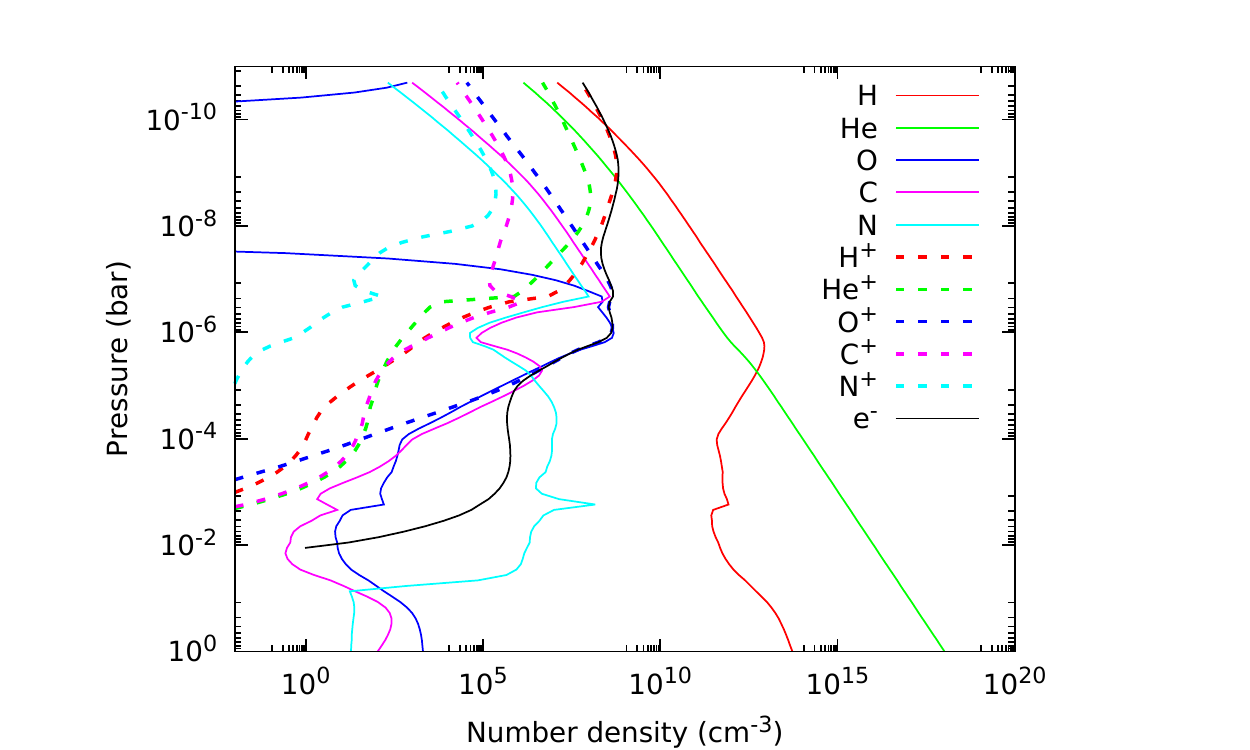} & \hspace{-2cm} \includegraphics[width=10cm]{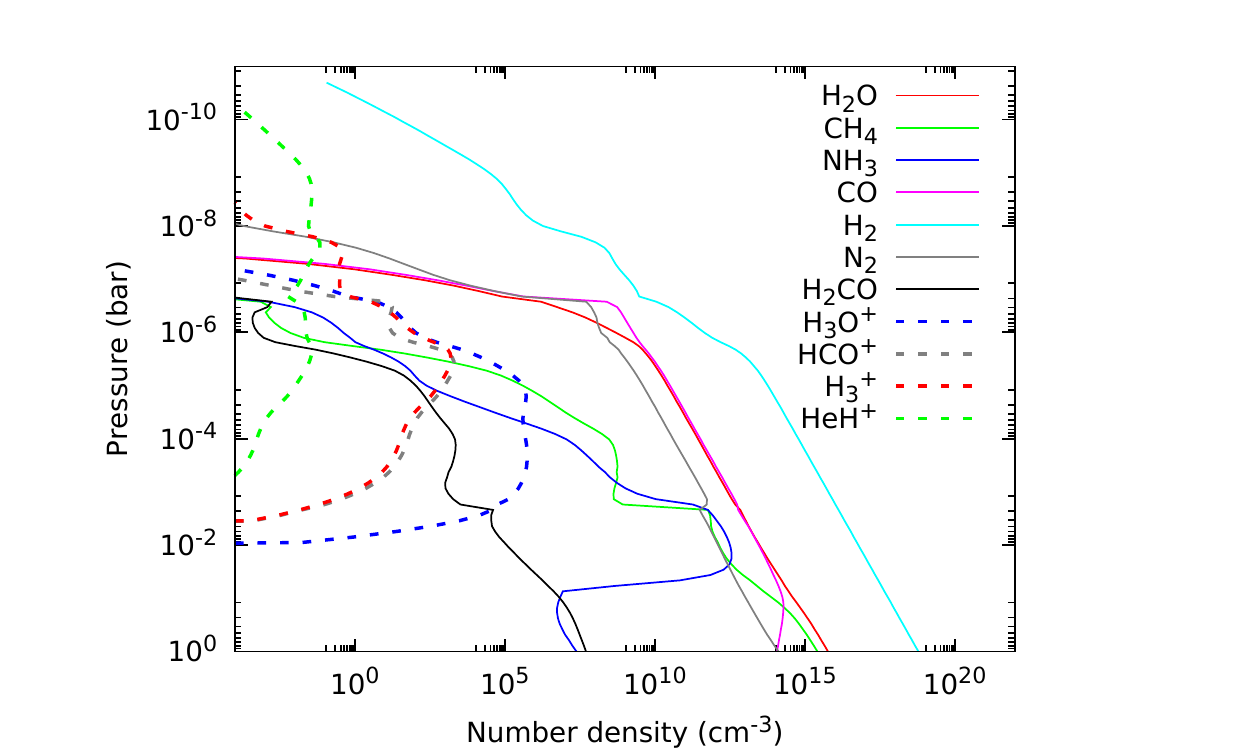}\\
\end{tabular}
\caption{Atomic (left panel) and molecular (right panel) vertical number density profiles for a selected set of species under the conditions of the RF model.}
\label{fsix}
\end{figure*}

\subsection{The spectral distribution of radiation within the atmosphere}
\label{tau}
In Figure \ref{ffive}a we show the resulting atmospheric optical depth at different photon energies. Being defined through equation (\ref{tauh}), the optical depth at a given energy is sensitive to the concentration of a chemical species, and its ability to interact with radiation in that energy range. In the topmost layers, the Lyman$-\alpha$ line central frequency is optically thin. This line turns into thick ($\tau \ga 1$) at $P \sim 5 \times 10^{-7}$~bar. When the pressure reaches the $10^{-5}$~bar level, the optical depth has increased of about 4 orders of magnitude, because of the rising densities of molecular species (see the upper plateau in the dissociation rates, Figure \ref{ffour}a). From this location on, Lyman$-\alpha$ radiation is virtually extinct, see Figure \ref{ffive}b in which it is shown the pressure at which a photon of a given energy reaches the apical value of $\tau = 100$. Higher energy photons e.g., $E =14$~eV a value slightly larger than the energy threshold of atomic hydrogen ionization, present optical depths increasing rapidly up to pressures at which molecular species begin to form, $P \sim 10^{-5}$~bar. There, most of hydrogen atoms are segregated in molecular compounds. We find that, in general, EUV radiation is totally removed starting from $P \ga 10^{-6}$~bar. Higher energy radiation is progressively damped with increasing pressure, however remaining still capable to maintain a chemically significant ionization level. Around $P \sim 10^{-4}$~bar, the gas (now predominantly molecular) is illuminated by X-rays with energies $E \ga 300$~eV, that are able to penetrate quite easily through the atmospheric layers, reaching out very low altitudes. As we already pointed out, since photoionization cross-sections scale with a negative power of the energy, the photon penetration depth increases with the energy. However, in the energy range exploited in this work, photons do not penetrate much lower than $0.1-1$~bar (depending on the illumination), where other phenomena, e.g., dynamics, may be dominant.

\subsection{The atomic and molecular distributions} \label{RMdistribution}
The atmospheric gas-phase abundances for our RF model are presented in Figure \ref{fsix}, for a selected number of atomic or molecular species, either in neutral and ionized forms.

As evident from Figure \ref{fsix} (left panel), recombination of atomic ions to neutrals occur rather high in the atmosphere for H and He, $P \la 10^{-10}$~bar, while C and N follow at a bit higher pressures $P \la 10^{-9}$~bar. Neutral oxygen reaches the concentration level of O$^+$ much more in depth, $P \sim 10^{-6}$~bar, as the $K-$shell ionization rate declines. These transition regions depend on the physical and chemical conditions of the atmosphere, as reported in Table \ref{tthree} in which we summarize the pressure crossing points between ions and neutrals for all the atoms present in the network. Together with the RF model, we report the crossing points for most of the models listed in Table~\ref{tone}. It is worthwhile to recall that in the NX model, the EUV spectral band is present, and identical to the one of the RF model. The first evidence stemming out from these results, is that at the topmost altitudes, ionizations are entirely driven by EUV radiation. In fact, no crossing point exists when EUV radiation is suppressed (NE model), and no deviations from the results of the RF model appear in the NX (no X-rays) and NS (no secondary effects) models. Oxygen is the exception, exhibiting a crossing point in the NE model, suggesting that, although EUV radiation is still the dominant driver of primary ionization, X-rays begin to contribute (see the NX model). Moreover, secondary ionization appears to be rather effective (NS model).
\begin{figure}
\centering
\includegraphics[width=9cm]{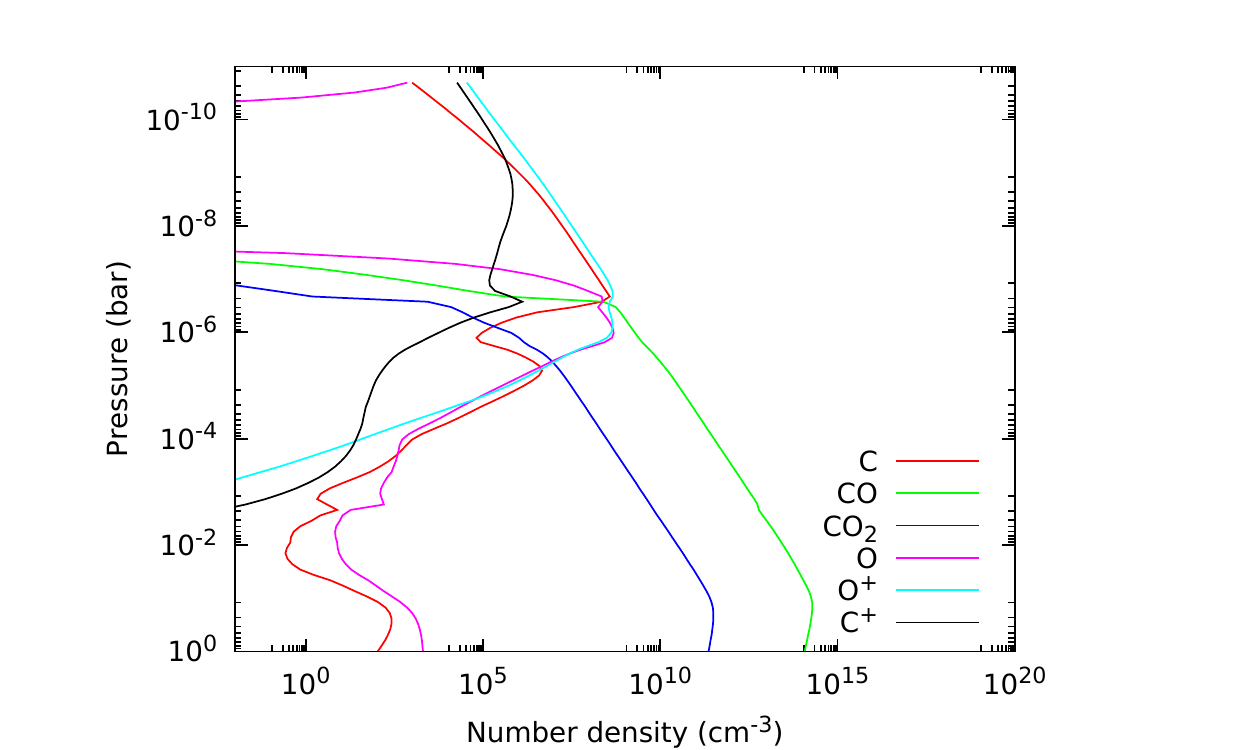}
\caption{The C$^+$/C/CO transition.}
\label{fseven}
\end{figure}

These trends are due to the different photon penetration depths with higher energy photons penetrating deeper into the atmosphere. Since the oxygen $K-$shell ionization threshold is located at larger energies than those of carbon ($\Delta E \sim 200$~eV) and nitrogen ($\Delta E \sim 100$~eV), ionization events are produced at lower altitudes, at which the EUV radiation has began to be significantly damped. As X-rays are extremely sensitive to  the presence of metals, variations in the metallicity  may impact on the atomic ions distribution. We find that, while C and N are scarcely affected, oxygen concentration responds quite appreciably to variations in the metallicity (LM and HM models). Increasing the stellar activity both EUV and X-ray band fluxes increase (together with that of the Lyman$-\alpha$). However, since their relation is almost linear, $\sim L_{\rm X}^{0.86}$ \citep{Sanz-Forcada11}, while the attenuation is exponential (see equation \ref{expo}), the increase in the activity extends the region dominated by X-rays far beyond that in which EUV radiation matters. The net effect is an increase of the X-ray influence over the ionization (HA model). We also note that some elements, e.g.,  oxygen may have additional deeper crossing points, due to the interplay of chemical reactions.

Recombination of hydrogen ends up in the formation of H$_2$ (the most abundant species), and at larger pressures also nitrogen becomes molecular. Once the concentrations of O$^+$ declines, major repositories of oxygen are carbon monoxide and water. In a limited range of pressure ($P \sim 10^{-9} - 10^{-6}$~bar) the neutral carbon concentration encompasses all carbon nuclei, giving rise to a well defined C$^+$/C/CO transition (Figure \ref{fseven}), characteristic of interstellar photodissociation regions (e.g., \citealt{Sternberg95}). In interstellar conditions (i.e. no radiation with energy higher than the Lyman continuum), however neutral atomic carbon is typically under abundant. In gas subjected to high energy photon irradiation, oxygen ionization is much more extended than that of carbon, and CO formation is delayed towards higher pressures, allowing the neutralization of carbon ions. At the same pressure level than CO, H$_2$O begins to form. Its formation rate is partially inhibited by the ionization content of the gas, while its destruction is mainly powered by UV radiation. Ammonia and methane increase their abundances deep in the atmosphere (Figure \ref{fsix}b), although in lower concentrations than CO and H$_2$O. 

As secondary ionization starts to dominate the electronic content, i.e. when photochemistry is mainly driven by X-rays, heavy molecular ions begin to form. These species are less abundant than neutral species, reside at mid-altitudes in the atmosphere ($P\sim 10^{-7} - 10^{-2}$~bar), and are mostly dependent on radiation chemistry. One of the most abundant is the hydronium ion, H$_3$O$^+$ resulting from the protonation of water. The absence of H$_2$O$^+$ suggests an active proton-transfer chemistry, $ \rm H_3^+ + H_2O \to H_3O^+ + H_2$. H$_3^+$ depends on the electron content, as its formation occurs through $\rm H_2 + H_2^+ \to H_3^+ + H$, and the dihydrogen cation, H$_2^+$ is formed mainly by electron impact ionization of H$_2$ (15.4~eV), at altitudes where EUV radiation is suppressed. H$_3^+$ is a universal protonator, initiating a chain of ion-neutral reactions that is responsible for the formation of many molecular ions, such as HCO$^+$, coming from protonation of CO. 

\begin{figure}
\centering
\includegraphics[width=9cm]{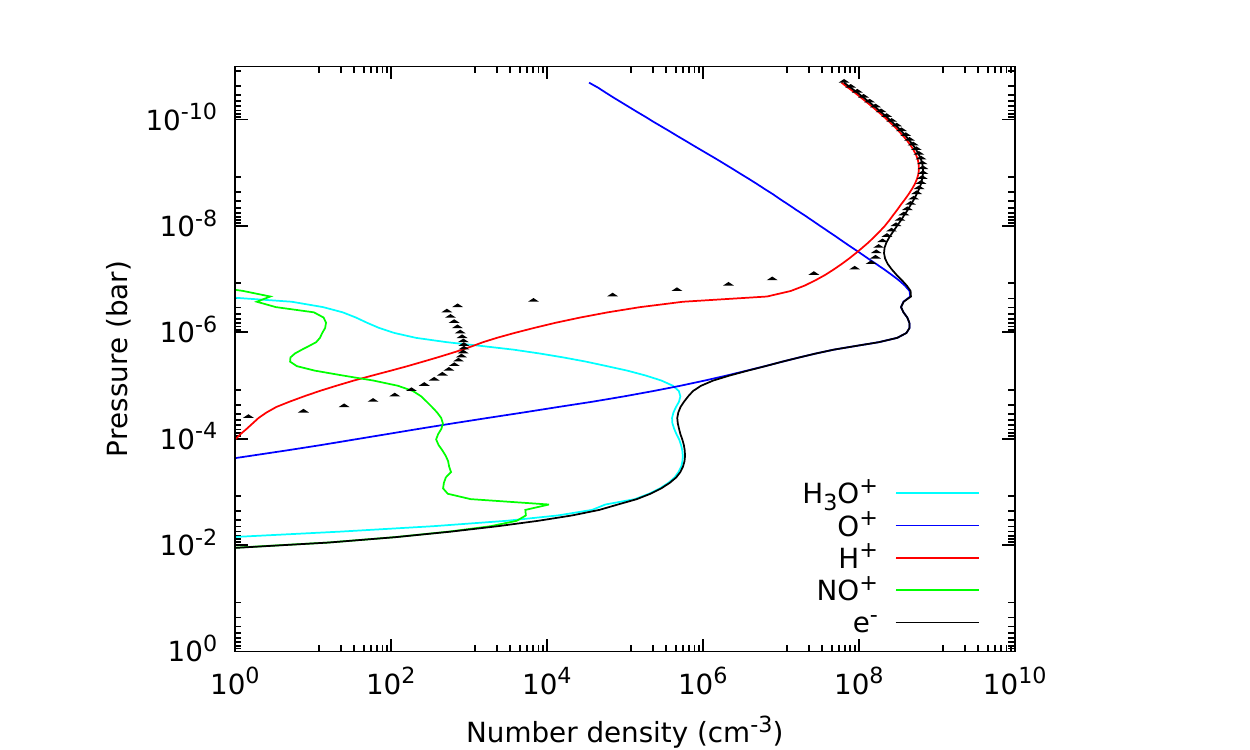}
\caption{Vertical density profiles of electrons and their major contributors under the conditions of the RF model (solid lines). Black dots indicate the distribution of the electron density suppressing X-rays (NX model).}
\label{feight}
\end{figure}
In Figure \ref{feight} we report the major indicators of the gas electron content. As discussed in the previous Section, with increasing pressures first hydrogen and then oxygen are the dominant providers of electrons. At lower altitude, the abundance profile of the hydronium ion (together with other metal molecular ions, e.g., NH$^+$) appears to be tightly correlated with the electron distribution, for reasons related to H$_3$O$^+$ formation channel (see last Section). 

\begin{figure}
\centering
\includegraphics[width=9cm]{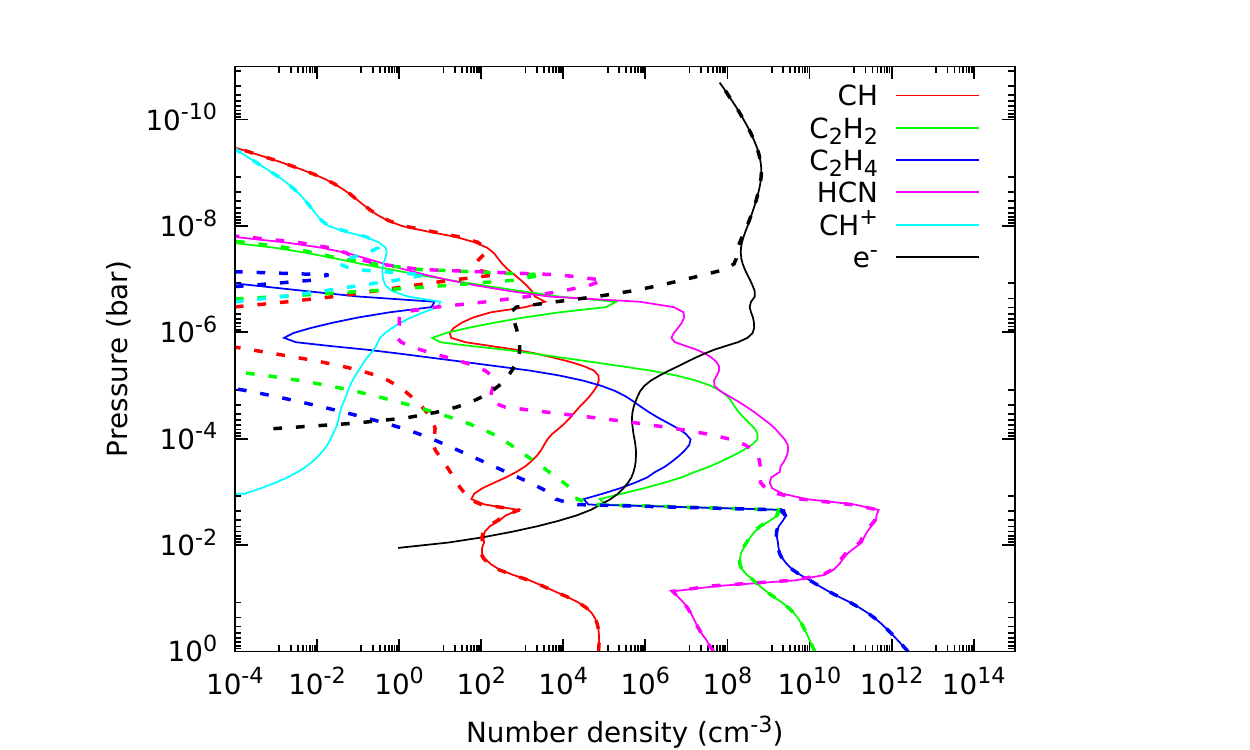}
\caption{The vertical distribution of some C-H bearing species including hydrogen cyanide (HCN) in the RF (solid line) and NX (dashed lines) models.}
\label{fnine}
\end{figure}

The methylidyne radical (CH) and cation (CH$^+$), and the hydrocarbons acetylene (C$_2$H$_2$) and ethylene (C$_2$H$_4$) show relatively large densities. Hydrogen cyanide (HCN) is also abundant. The concentration profiles of these species are riported in Figure \ref{fnine}. All of these molecules are positively sensitive to X-ray radiation. without which their abundances would to be basically confined towards the bottom of the atmosphere. Photochemistry boosts their abundances upwards at altitudes at which haze formation is expected, $P \sim 10^{-2} - 10^{-3}$~bar \citep{Madhusudhan19}. 

\subsection{Varying the stellar activity} \label{XV}
As we have seen, chemistry may be sensitive to the effects of the different types of high-energy radiation. The characteristics of atmospheric chemical evolution emerge from many feedbacks on a wide range of time scales. Identifying and quantifying these processes is an essential step in the predictive outcome of our chemical modelling.  

\begin{figure*}
\centering
\begin{tabular}{ccc}
\includegraphics[width=6cm]{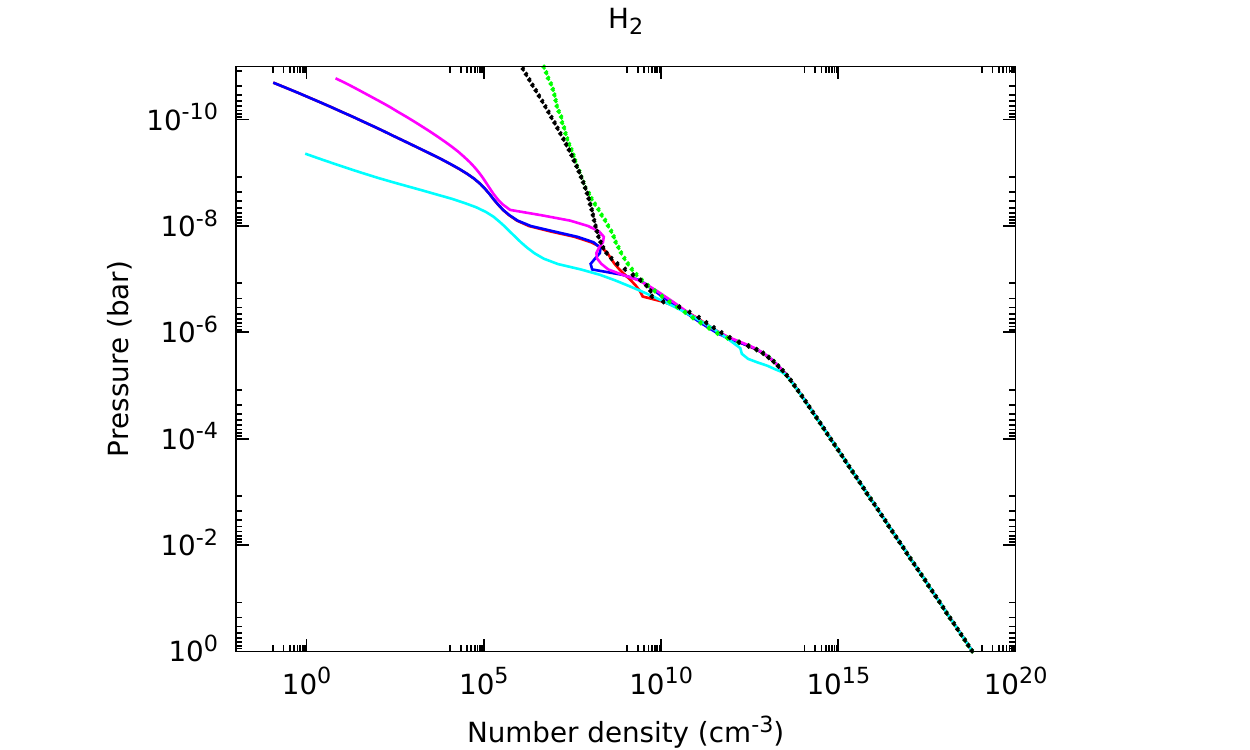} &  \includegraphics[width=6cm]{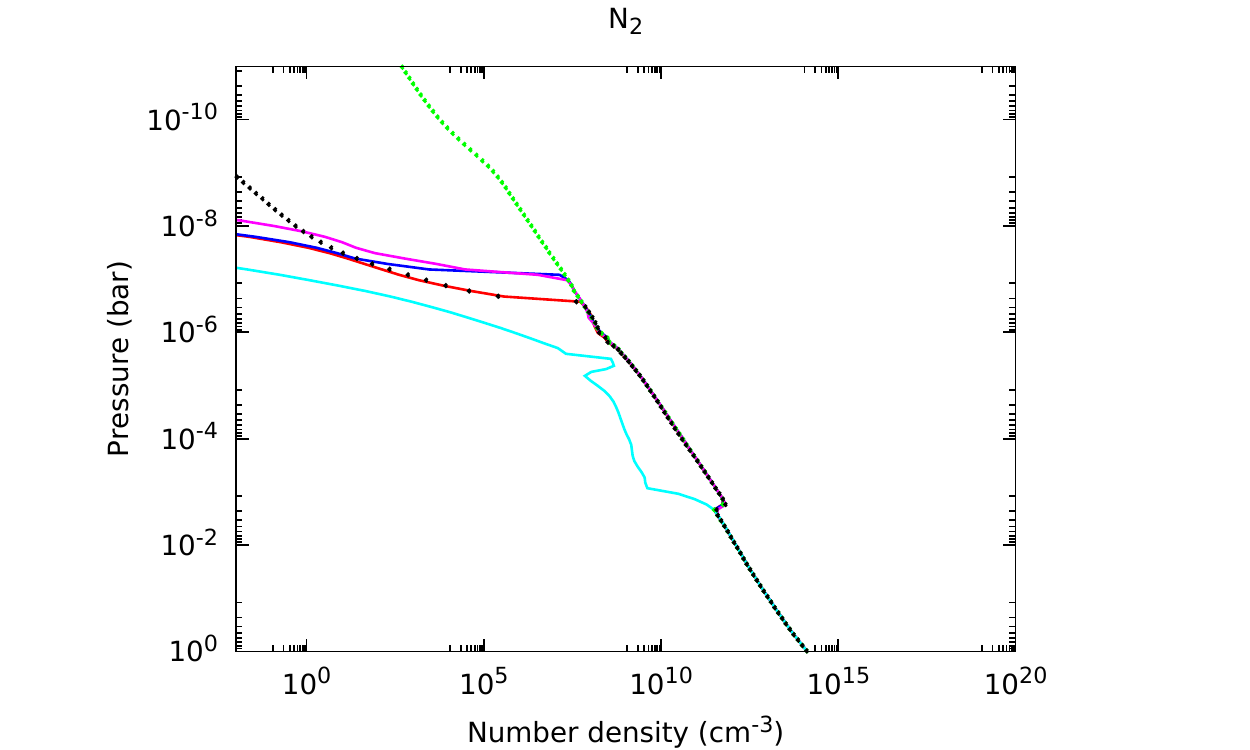} &  \includegraphics[width=6cm]{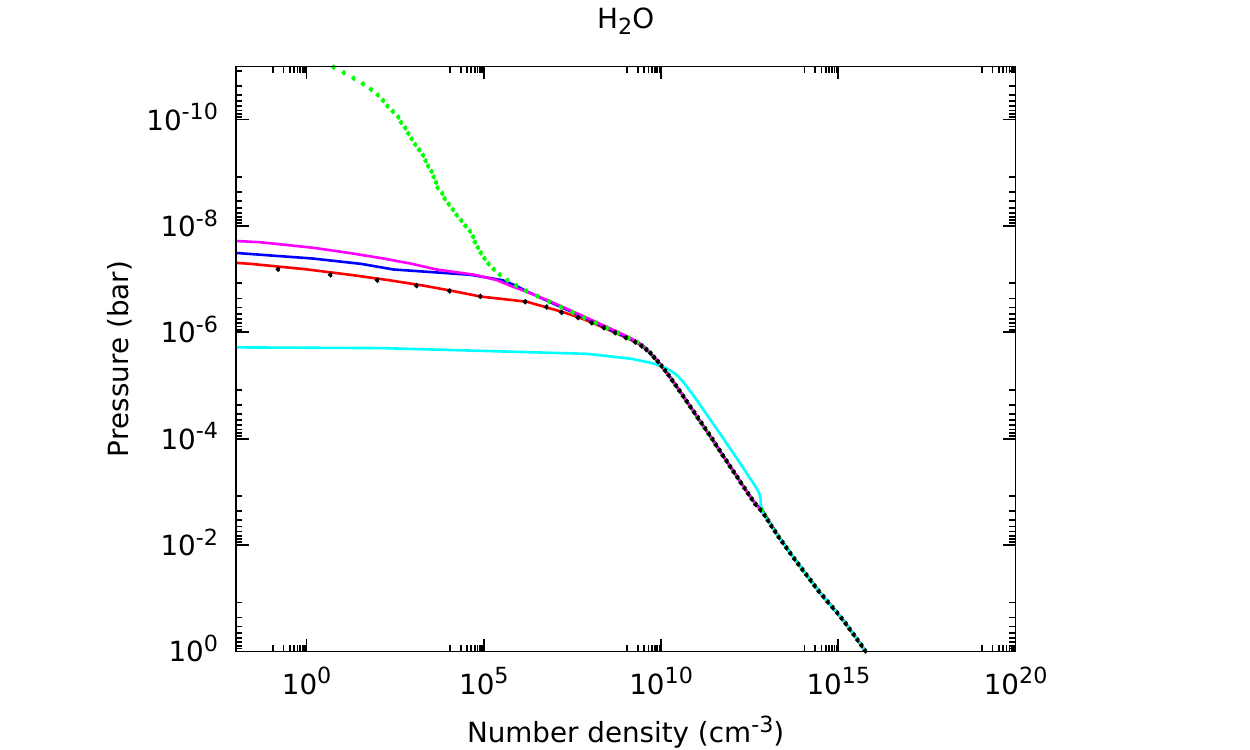} \\
\includegraphics[width=6cm]{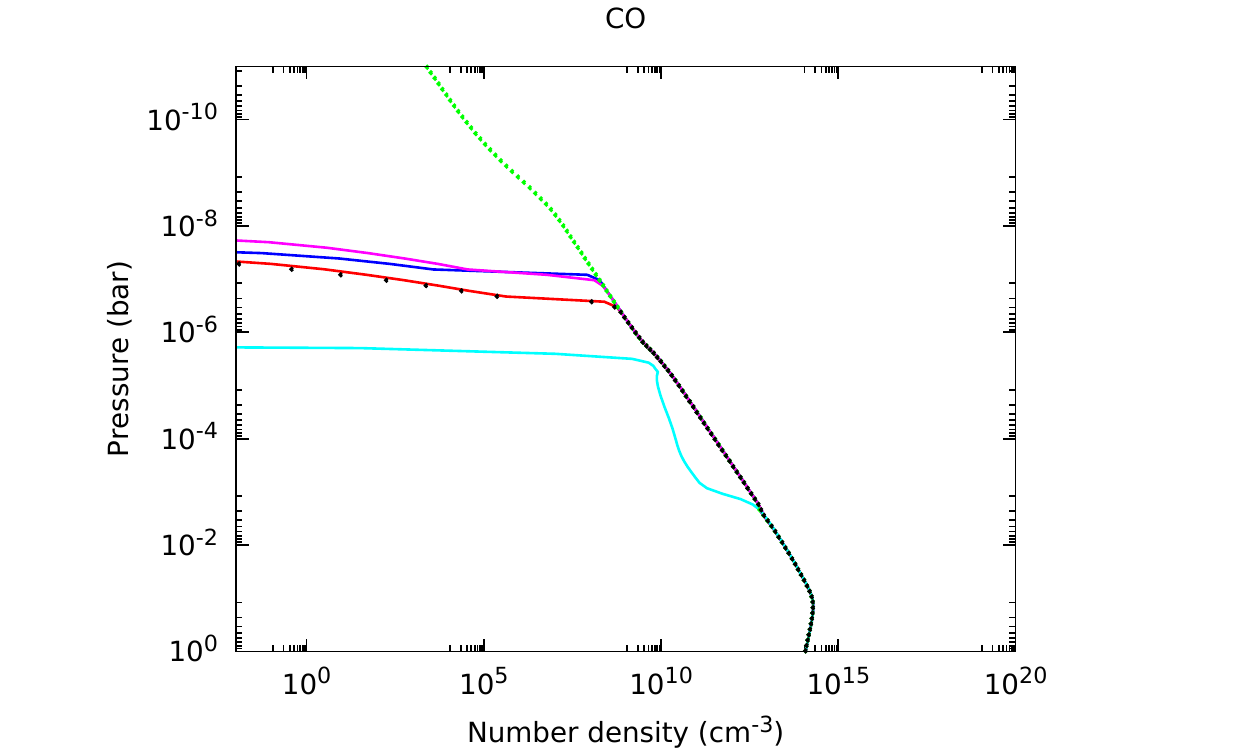} &
\includegraphics[width=6cm]{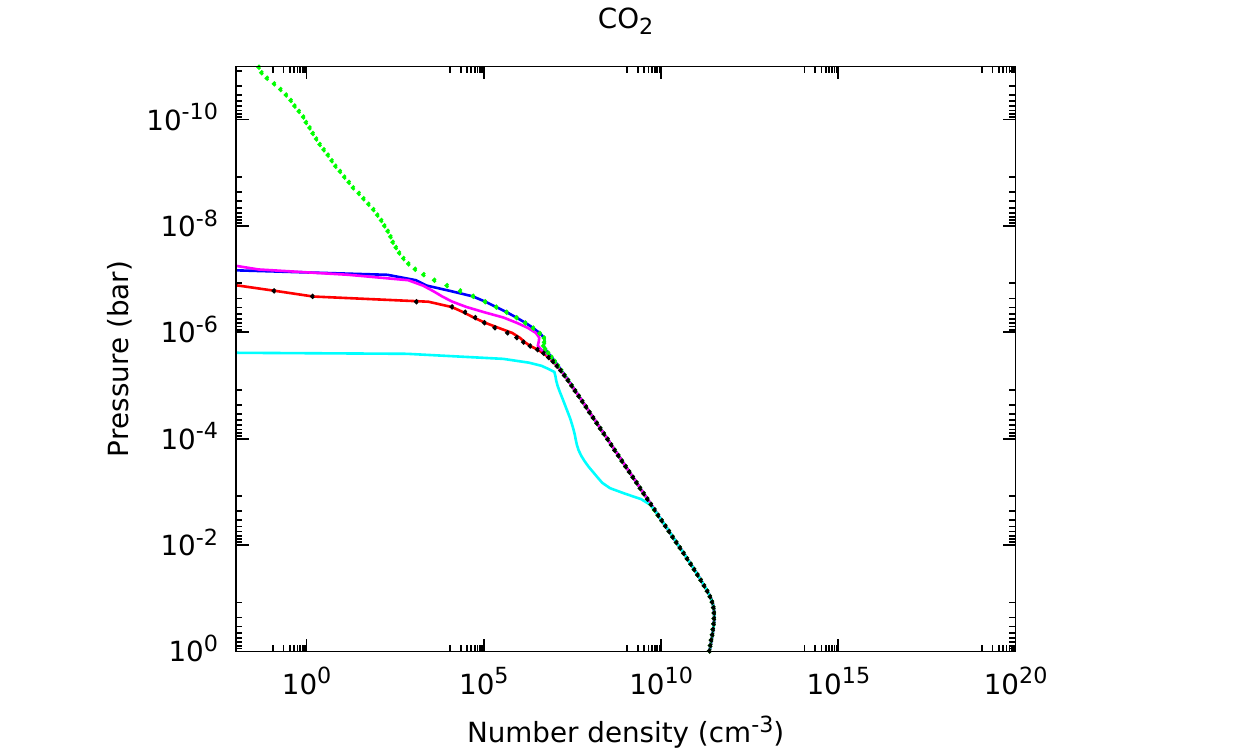} &
\includegraphics[width=6cm]{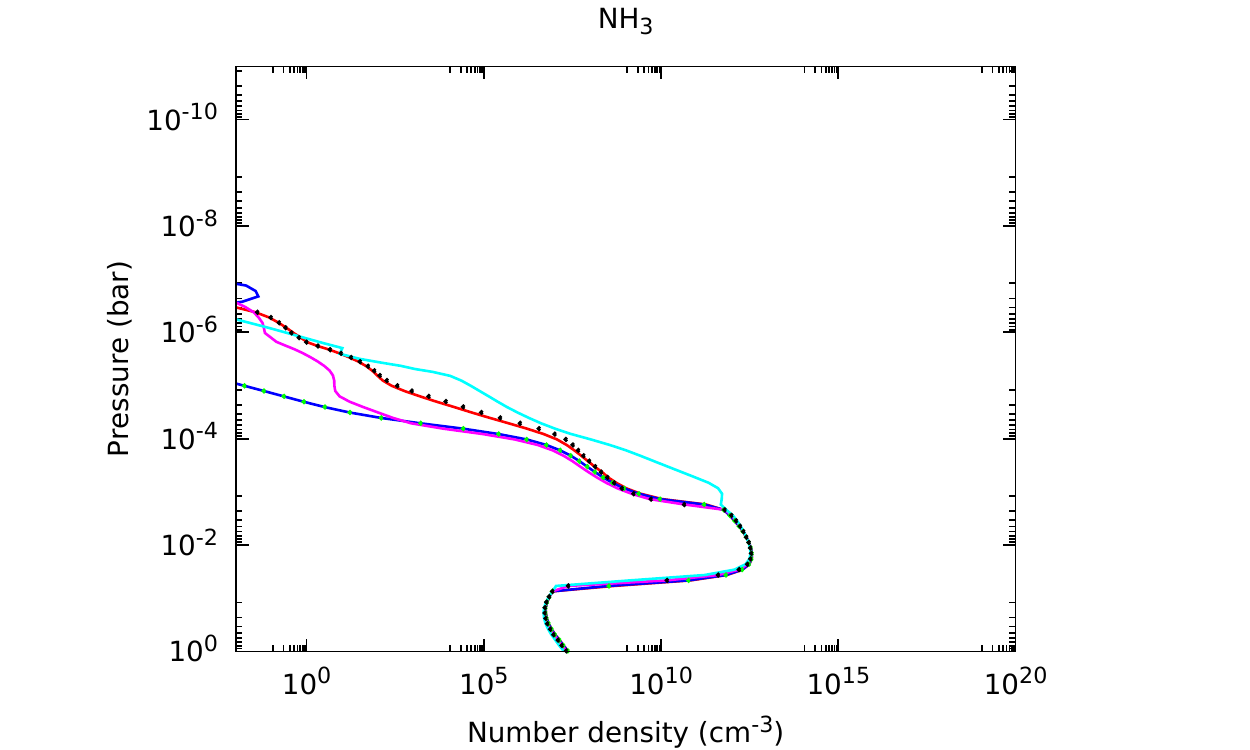} \\
\includegraphics[width=6cm]{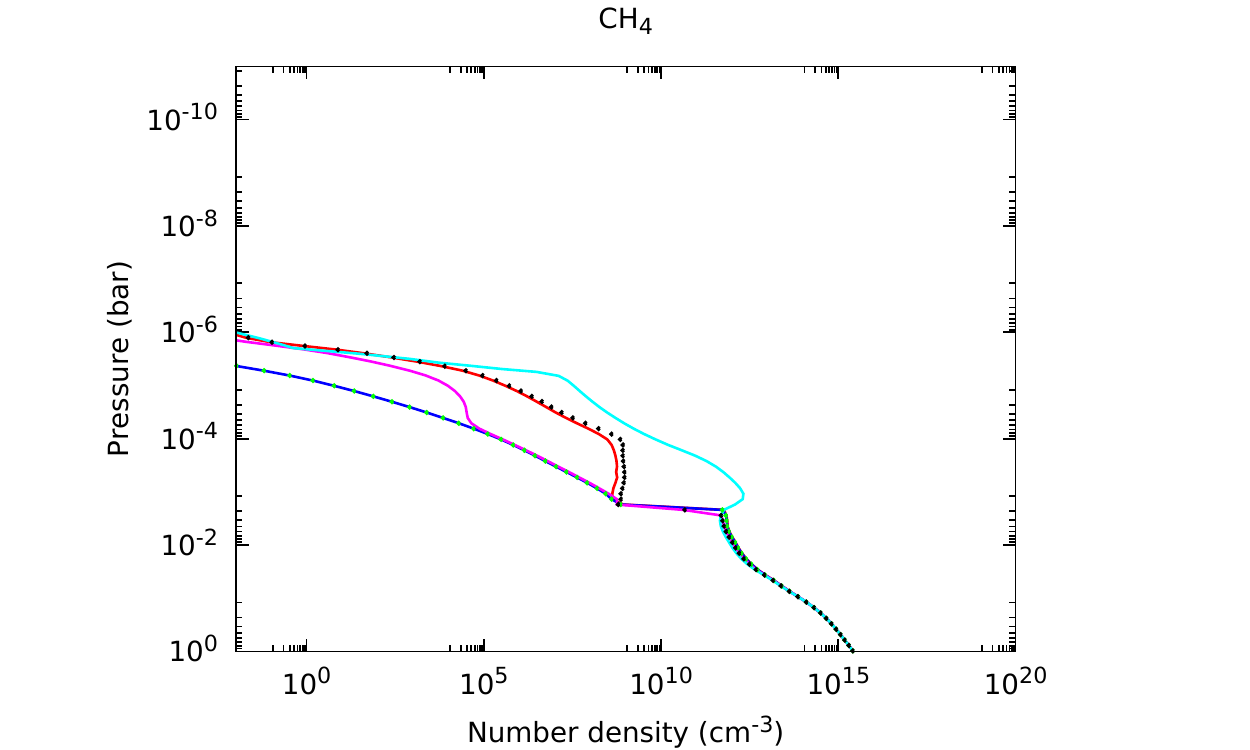} &  \includegraphics[width=6cm]{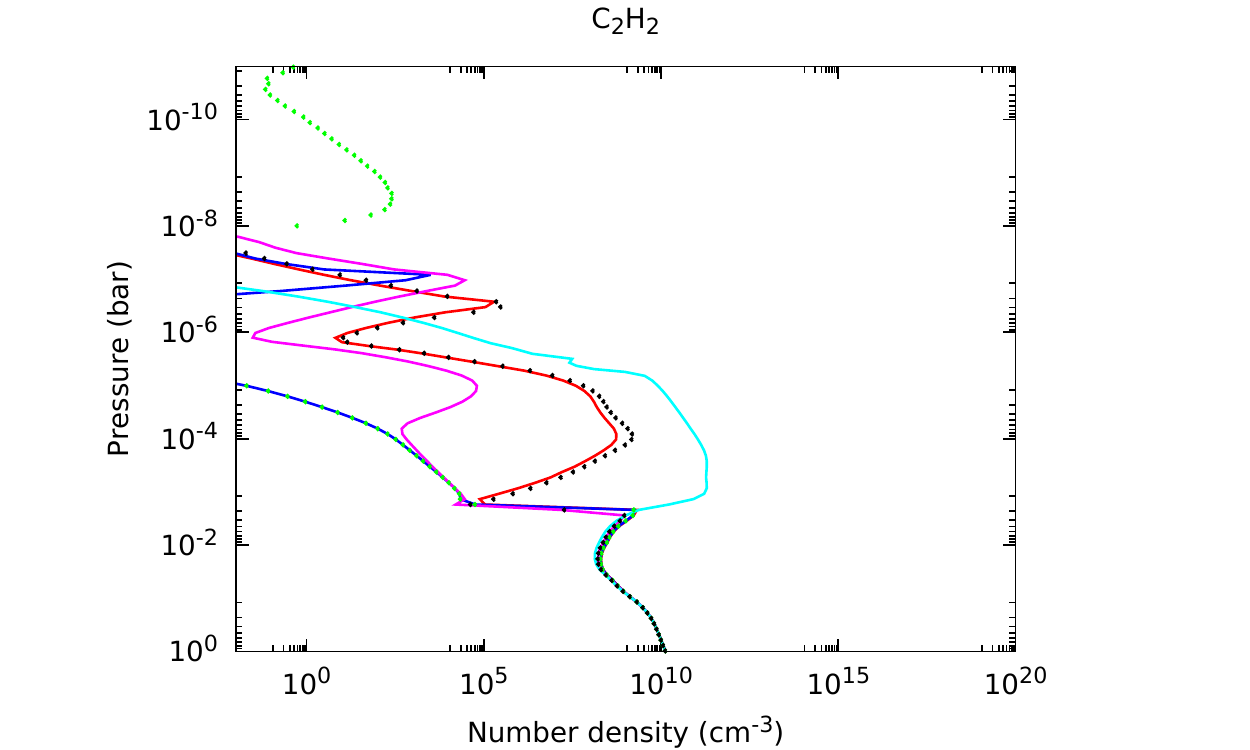} & \includegraphics[width=6cm]{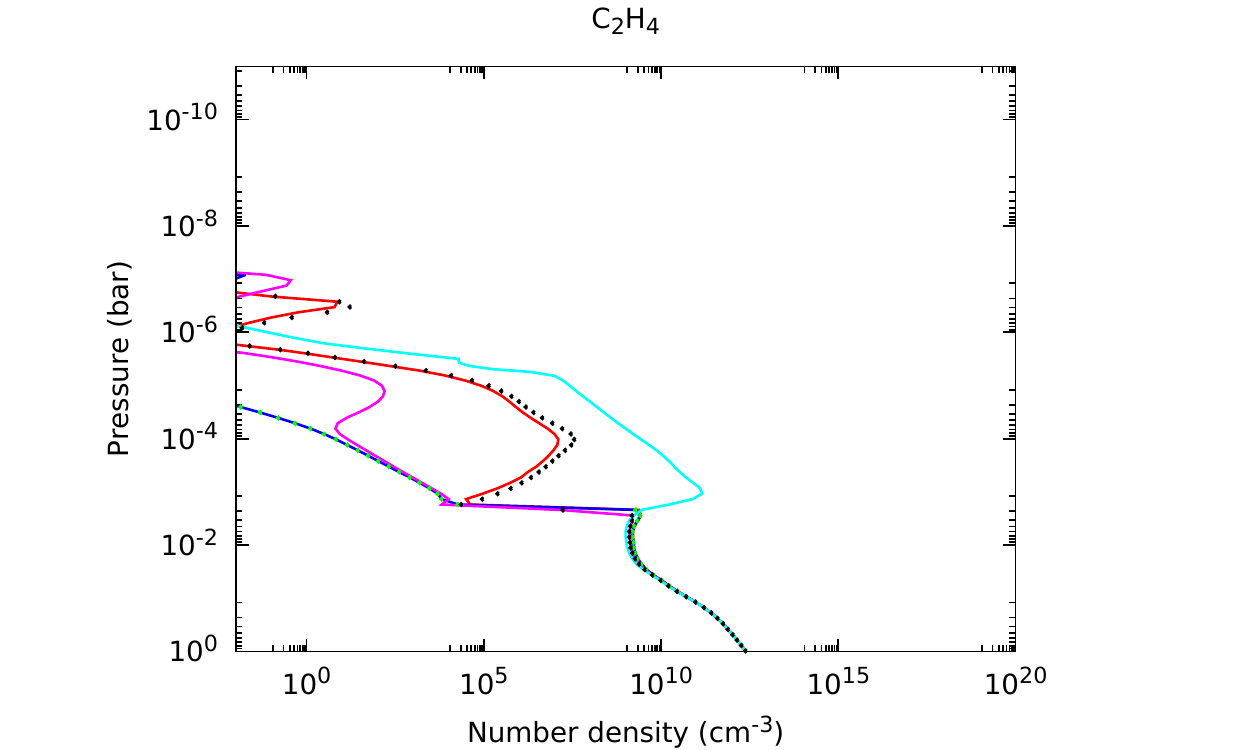} \\
\includegraphics[width=6cm]{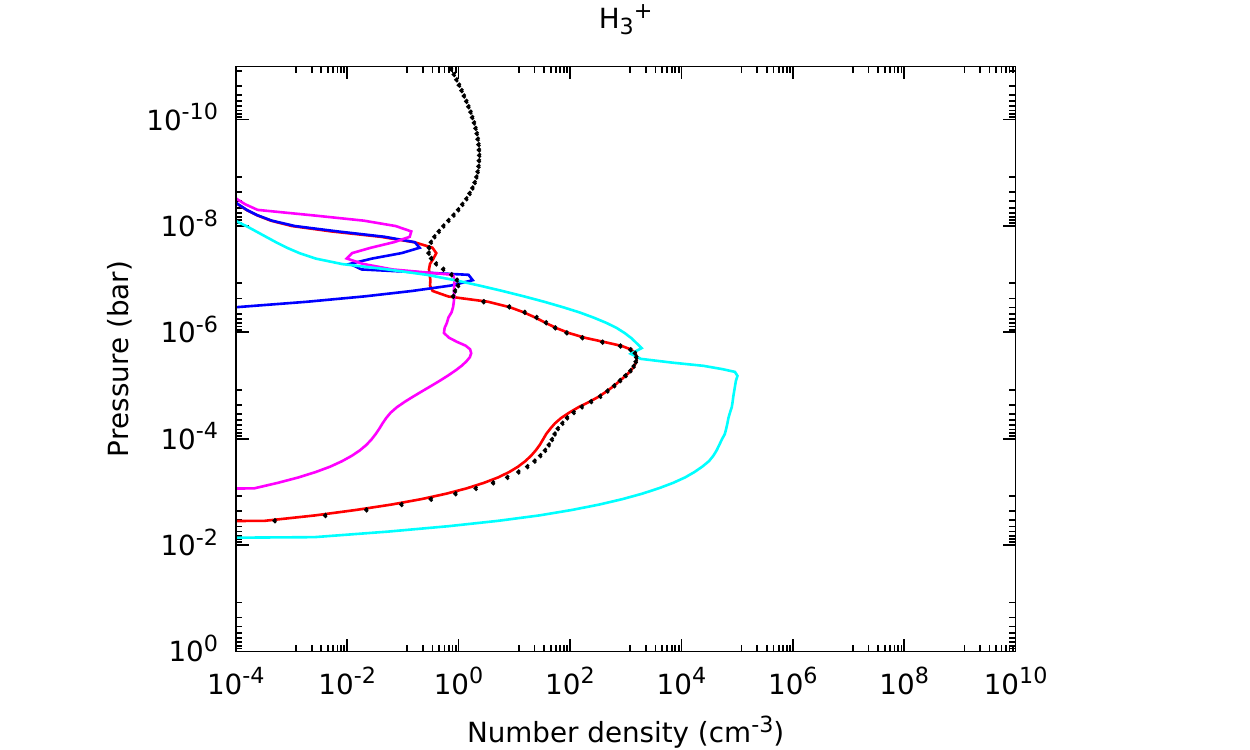} &
\includegraphics[width=6cm]{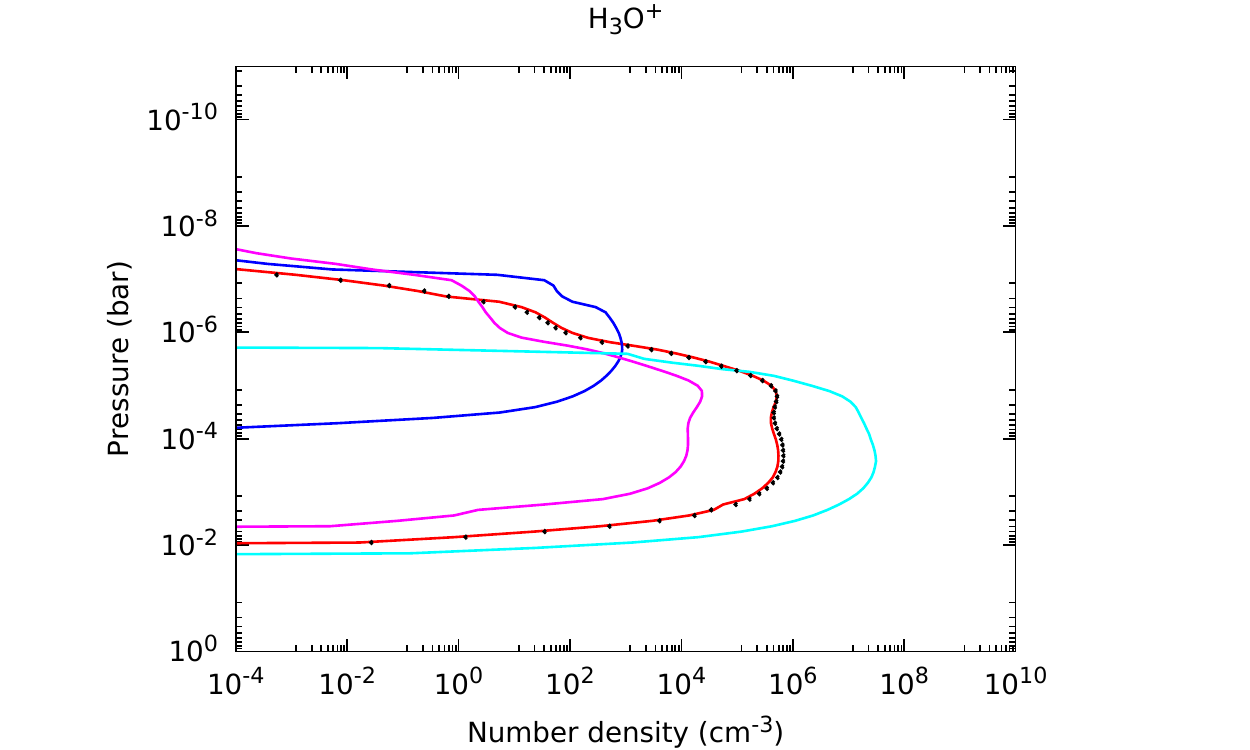} &
\includegraphics[width=6cm]{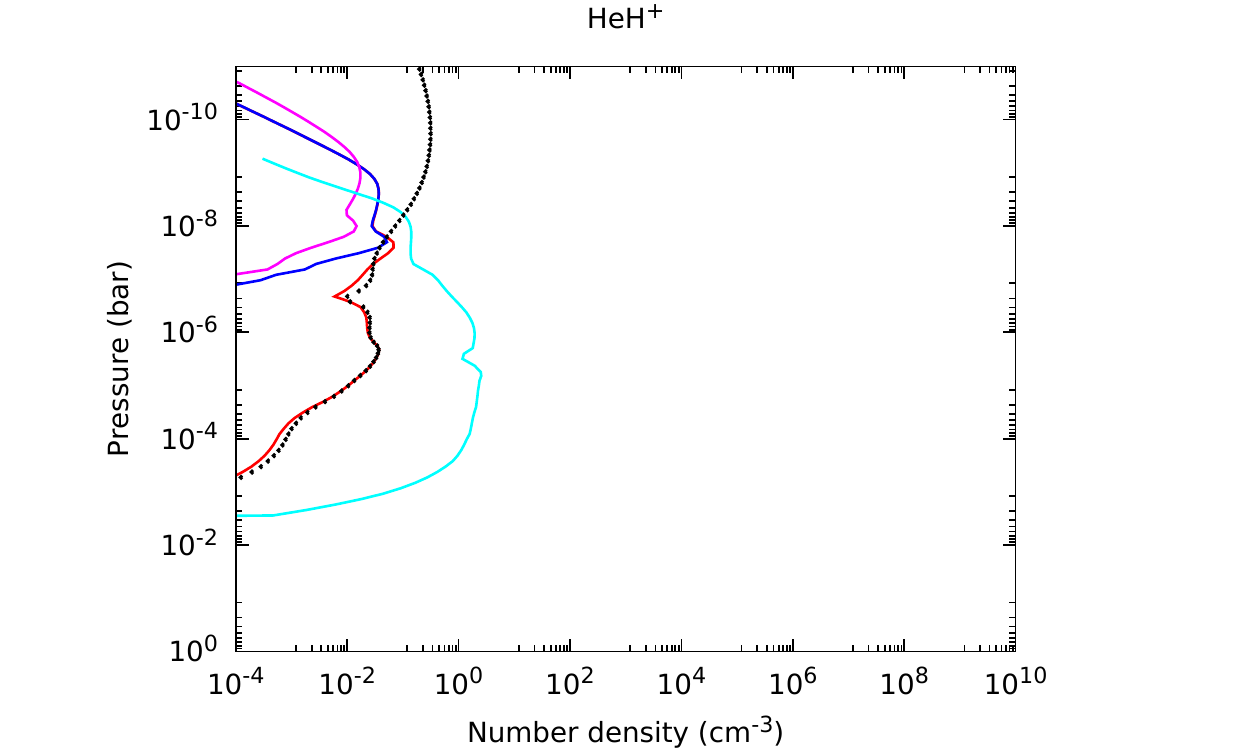}
\end{tabular}
\caption{Atmospheric concentrations arising from different assumptions on the illuminating stellar high energy flux. RF model ($L_{\rm X} = 1 \times 10^{28}$~erg~s$^{-1}$, $T_{\rm X} = 0.5$~keV): red lines; LA model ($L_{\rm X} = 1 \times 10^{27}$~erg~s$^{-1}$, $T_{\rm X} = 0.3$~keV): purple lines; HA model ($L_{\rm X} = 1 \times 10^{30}$~erg~s$^{-1}$, $T_{\rm X} = 1$~keV): pale blue lines; NX model (no X-rays): blue lines; UV model (no radiation with energy higher than the Lyman continuum): green dots; NE model (no EUV radiation): black dots.}
\label{ften}
\end{figure*}

In Figure \ref{ften} we show the modifications in the abundances of a set of representative species, in response to different assumptions on the intensity of the stellar ionizing radiation, i.e. in our context, the radiation of energy higher than the Lyman continuum.
We find a clear separation in the response of molecular vertical profiles yielding two distinct trends, some species having their abundances reduced by the increase in the stellar activity, while others resulting enhanced. In the first group we find (among others) H$_2$, N$_2$, water, carbon monoxide, and CO$_2$. We recall that we parametrize the stellar activity through the X-ray luminosity, which in turn determines the intensity in the EUV spectral band and Lyman$-\alpha$ line. Thus, it may happen that for some species the X-ray luminosity acts just as a tuning parameter, e.g., enhancing the EUV intensity, but without effectively contributing to the chemistry. This is the case for molecular hydrogen, whose response to a different degree of stellar activity is driven entirely by EUV radiation, being practically insensitive to X-ray irradiation. This is easily understood on the base of the values assumed by the H$_2$ photoionization cross-section ($\gamma \sim 3.2$) at the thresholds of EUV and X-ray spectral bands
\begin{equation}
\frac {\sigma_{\rm H_2} (13.6 \, {\rm eV})} {\sigma_{\rm H_2} (100 \, {\rm eV})} \sim \left( \frac {100}{13.6} \right)^{3.2} \sim 600
\end{equation}
However, deeper in the atmosphere, H$_2$ may present a residual ionization produced by collisions with the secondary electron cascade generated by X-rays.

The dependence of N$_2$, water, carbon monoxide and dioxide abundances on the X-ray flux is instead real, as traced by the overlapping of their vertical profiles in both the RF and NE models. While CO is scarcely affected by UV radiation (UV model), water and CO$_2$ \citep{Venot18} may be efficiently dissociated by UV radiation, in particular in the energy range close to the Lyman-$\alpha$ line because of its great relative intensity. This is evidenced by the fall of H$_2$O and CO$_2$ profiles for pressure lower than $\sim 10^{-6}$~bar in UV and NX models. Due to the larger concentration of O$^+$, both water and carbon monoxide formation rates reduce with increasing ionization, and indeed they decline with increasing stellar activity, see the drop of CO and H$_2$O densities at pressures as large as $P \ga 10^{-5}$~bar in the high activity case (HA model). Moreover, the density profile of water does not show any change in the slope of its horizontal part, as compared to atmospheres with lower X-ray illuminations (LA and RF models). This means that the reduced water content is due to a shortage in reactants, rather than direct Lyman-$\alpha$ photodestruction. Even if in our model the Lyman$-\alpha$ intensity grows with the X-ray luminosity, the horizontal plateau in the abundance of water occurs at altitudes sufficiently low to make the UV radiation efficiently shielded (see Section \ref{tau}). The CO formation rate is affected by X-rays similarly to water. 

For the remaining neutral species and the molecular ions, the X-ray luminosity boosts their abundances as their formation channels increase with the electron content, or the products of dissociation. UV and EUV radiation play minor roles, as evidenced by the overlap of molecular profiles in the outcomes of NX and UV, and RF and NE models, respectively. Because of ionization, ammonia and methane benefit from additional sources of NH$_2$ and CH$_3$, respectively. This occurs through chains of hydrogen abstraction, that starting from NH$^+$ and CH$^+$, end up to form NH$_3^+$ and CH$_3^+$. At this stage, the chains are broken by electron dissociative recombination. 

H$_3^+$ and HeH$^+$ show different trends in the upper atmospheric regions, where EUV radiation provides an additional exit channel through electron dissociative recombination. As already mentioned, H$_2$ electron impact ionization, whose rate increases with the X-ray luminosity, provides a significant source of H$_2^+$ and then H$_3^+$ ions. Other molecular ions benefit on some level from electron impact chemistry (see HA model). 

\subsection{The role of metallicity}
Metallicity affects the chemistry in a relatively simple way, as  molecular abundances scale with the mutual amount of the elements. However, since the gas photoionization cross-section depends significantly upon the presence and the concentrations of heavy elements, the degree of ionization is affected by variations in the elemental composition, yielding consequences in the molecular distribution.  

\begin{figure}
\centering
\includegraphics[width=10cm]{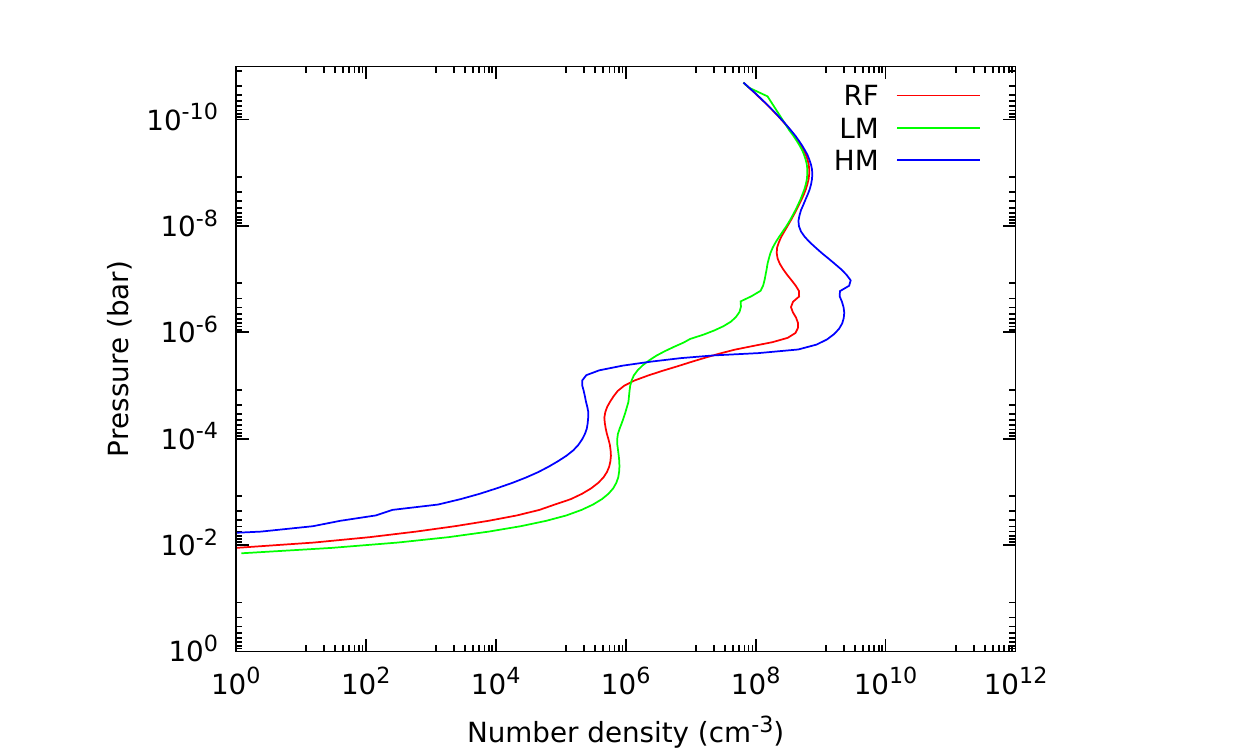}
\caption{Metallicity-dependent electron density profiles in the atmosphere. RF model: red line; LM model: green line; HM model: blue line.}
\label{feleven}
\end{figure}

In Figure \ref{feleven} we report vertical profiles of the electron number density for three representative values of the metallicity, namely the RF model,  $Z/Z_\odot = 0.1$ (LM model), and $Z/Z_\odot = 10$ (HM model). The outer atmospheric layers do not show appreciable variations, as most of electrons are provided by the photoionization of hydrogen. Since X-rays are more rapidly removed in the HM case, from $P \sim 10^{-8}$~bar down to $P \sim 10^{-5}$~bar the ionization initially increases with increasing metallicity, until the X-ray intensity declines appreciably, and the electron content falls sharply. The horizontal plateau in the electron density at $P \ga 10^{-6}$~bar (RF and HM models) is in fact, the response to the exponential decrease of the flux (see equation \ref{expo}), while formation of H$_3$O$^+$ provides a new source of electrons that temporarily brakes the fall of the ionization content of the gas down to $P \sim 10^{-3}$~bar (see next Section). In the LM case, for opposite reasons, facilitated X-ray penetration favours the increase in the electron density at lower altitudes, making their vertical density profile much smoother.

\begin{figure*}
\centering
\begin{tabular}{cc}
\hspace{-1cm}
\includegraphics[width=10cm]{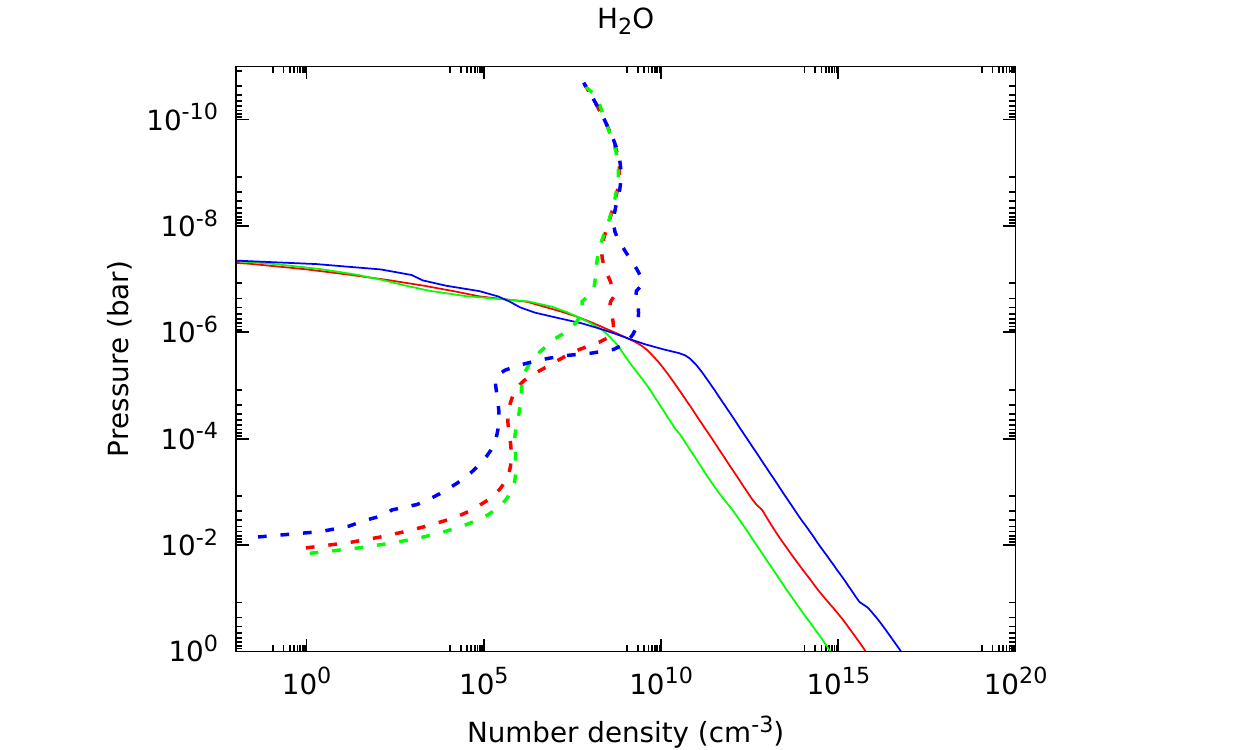} & \hspace{-2cm} \includegraphics[width=10cm]{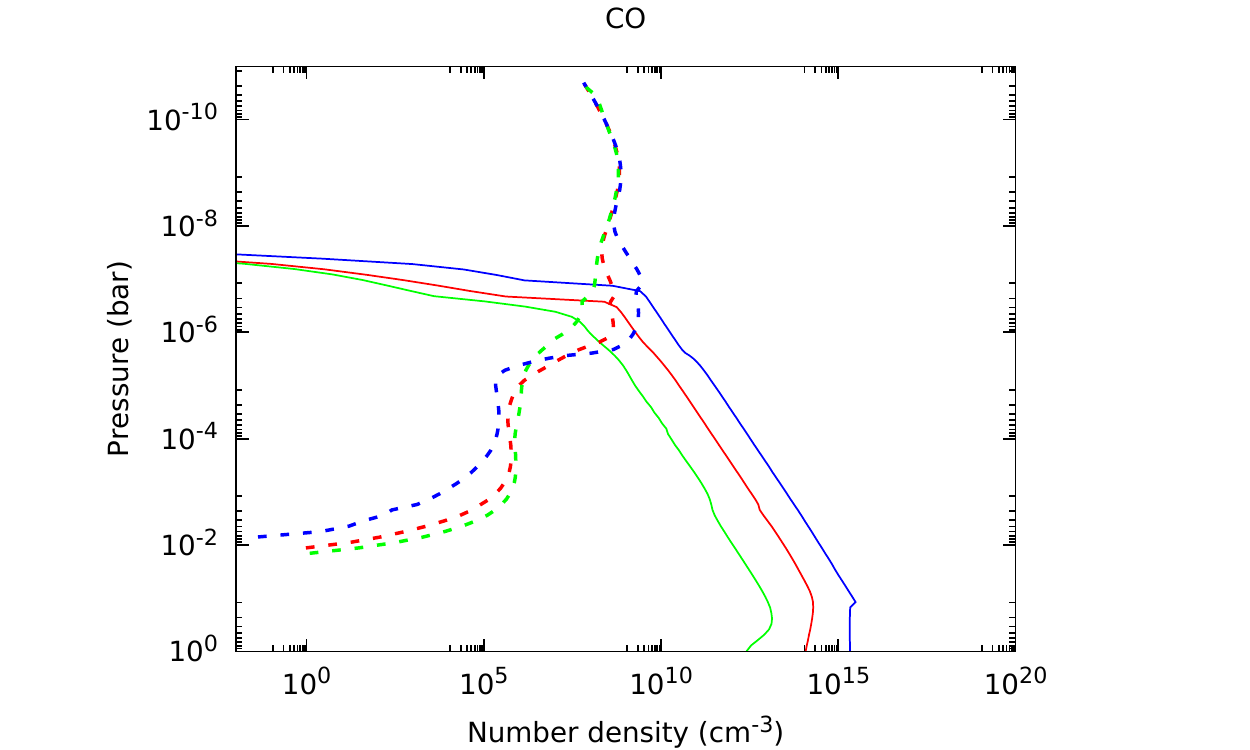}\\
\end{tabular}
\caption{Water (left) and carbon monoxide (right) vertical profiles. LM model: green lines; RF model: red lines; HM model: blue lines. Dashed lines refer to the electron profiles; colors indicate the same models as in the case of molecular species.}
\label{ftwelve}
\end{figure*}

Neutral species are affected at various degrees. For instance, the abundances of water are marginally perturbed, while CO and CO$_2$ may delay their formation up a few orders of magnitude in a small range of pressures (Figure \ref{ftwelve}). While the fall in the water abundance is started by UV radiation (mainly Lyman-$\alpha$), the coincidence of the electron and CO densities, at the change of slope of the CO profiles, suggests that X-rays are responsible for CO removal. This simply reflects the decay of oxygen ionization due to X-rays, as it is apparent by the overlapping of the profiles of O$^+$ and electrons (see Figure \ref{feight}). In other words, increasing the altitude CO is not destroyed, but stops to form efficiently. Since metallicity affects the penetration of ionizing radiation, the abundances of CO differentiate significantly in the three cases shown in Figure~\ref{ftwelve}, due to the shift of the electron, and thus O$^+$ abundance profiles towards lower values ($P \sim 10^{-7}$~bar, see Figure \ref{feleven}).

\section{Discussion and conclusions}
In this work we present an analysis of chemical processes induced by high energy radiation, in a planetary atmosphere of solar-like composition. In particular, one of the major effort in our investigation is the attempt to unfold the effects carried out by photons in different spectral bands. These effects frequently appear entangled and mixed by the interplay of chemical reactions, that provide a convolved response to radiation in the resulting molecular abundances. Such a quest is made even more difficult by the existing  correlations among different portions of the stellar spectrum. 

The spectral distribution of radiation within the atmosphere reflects the underlying photochemistry. While EUV radiation sets up the chemical (mainly atomic) distribution in the upper atmospheric layers, interacting predominantly with hydrogen and helium bearing species, its driving role tends to fade in favour of X-rays, more sensitive to the presence of heavy elements. We may place such a "boundary" at the location where carbon and oxygen are eventually bound into carbon monoxide molecules, i.e. at pressures (RF model) $P \sim 10^{-7}$~bar. Of course, no unique and sharp separation exists between EUV and X-rays dominated regions, and their mutual extent and overlapping depend on the elements involved, reaction rates, and the physical and boundary conditions. 

The major conclusion of the present work is that X-rays are a fundamental ingredient in the chemistry of planetary atmospheres of gaseous giants. X-rays with their weak photoionization cross-sections may push the gas ionization to pressures inaccessible to lower energy radiation. Although X-rays interact preferentially with metals, the produced secondary electron cascade may collisionally ionize also hydrogen and helium bearing species, and this occurs at altitudes far below the ones UV and EUV photons may penetrate. 

X-ray irradiation  supplies molecular ions that give potentially observable signatures of the atmospheric ionization. A specific example is H$_3$O$^+$, produced through protonation of the water molecules. The presence of abundant stable species inhibits routes involving hydrogen abstraction chains, such the one related to hydronium ion, initiated by the formation of the hydroxyl cation OH$^+$, which can react with molecular hydrogen to form H$_2$O$^+$ and then H$_3$O$^+$. As the major formation channel is $\rm H_2O + H_3^+ \to H_3O^+ + H$, the abundance of the hydronium ion is related to the one of H$_3^+$, which in turn depends on that of H$_2^+$, produced by the electron impact ionization of H$_2$. As a consequence, for each hydronium ion, one electron is produced. Since ionization of H$_2$ at those altitudes is the main source of electrons, this explain the tight correlation between the abundance profile of H$_3$O$^+$ and the electron concentration.

Chemical effects are not solely benign, as strong X-ray irradiation lowers appreciably the upper boundary of the residing regions of abundant species, such as water, CO and CO$_2$ (from $P \sim 10^{-7}$, RF model to $\sim 10^{-5}$~bar, HA model). At the same time however, ammonia and methane increase their concentrations. The same occurs for hydrocarbons and HCN, that constitute the chemical base for photochemically generated hazes. Such hazes are of particular interest as they may serve as a source of organic materials for potential chemical evolution of life on a planet \citep{He20}. Although life is not certainly expected to originate in giant hot planets, this chemistry may provide information on the relevant reaction routes. 

In conclusion, we have shown that stellar high energy emission, in particular X-rays, may drives important changes in the mixing ratio profiles of atmospheric species. The strongest impact on the chemistry is expected in planets orbiting stars of young ages that have the highest level of chromospheric activity. However, as the lowest adopted value of X-ray luminosity $L_{\rm X} = 1 \times 10^{27}$~erg~s$^{-1}$ is typical of old stars (e.g., \citealt{Schmitt04}), all the planets within a distance from the star $\la 0.05$~AU are affected by EUV and X-ray stellar radiation during their entire life. 

Finally, we note that stars are variable in time, and they may be subjected to flares and other impulsive phenomena that rise their emissions for a limited amount of time. These periods of high activity may increase photochemical and ionization rates, and thus impact atmospheric chemistry (e.g., \citealt{Venot16}), and even provide persistence in the products of chemistry \citep{Chen21}. The present analysis needs thus, to be extended to erratic stellar emission. 

We acknowledge contributions from ASI-INAF agreements 2021-5-HH.0 and 2018-16-HH.0. AM acknowledges partial support from PRIN INAF 2019 (HOT-ATMOS). We would like to thank the anonymous referees for their comments that helped to improve the clarity of the manuscript.

\bibliography{chemexoR1.bib}{}

\begin{thebibliography}{}
\expandafter\ifx\csname natexlab\endcsname\relax\def\natexlab#1{#1}\fi
\providecommand{\url}[1]{\href{#1}{#1}}
\providecommand{\dodoi}[1]{doi:~\href{http://doi.org/#1}{\nolinkurl{#1}}}
\providecommand{\doeprint}[1]{\href{http://ascl.net/#1}{\nolinkurl{http://ascl.net/#1}}}
\providecommand{\doarXiv}[1]{\href{https://arxiv.org/abs/#1}{\nolinkurl{https://arxiv.org/abs/#1}}}

\bibitem[{{{\'A}d{\'a}mkovics} {et~al.}(2011){{\'A}d{\'a}mkovics}, {Glassgold},
  \& {Meijerink}}]{Adam11}
{{\'A}d{\'a}mkovics}, M., {Glassgold}, A.~E., \& {Meijerink}, R. 2011, \apj,
  736, 143, \dodoi{10.1088/0004-637X/736/2/143}

\bibitem[{{Ag{\'u}ndez} {et~al.}(2014){Ag{\'u}ndez}, {Parmentier}, {Venot},
  {Hersant}, \& {Selsis}}]{Agundez14}
{Ag{\'u}ndez}, M., {Parmentier}, V., {Venot}, O., {Hersant}, F., \& {Selsis},
  F. 2014, \aap, 564, A73, \dodoi{10.1051/0004-6361/201322895}

\bibitem[{{Airapetian} {et~al.}(2016){Airapetian}, {Glocer}, {Gronoff},
  {H{\'e}brard}, \& {Danchi}}]{Airapetian16}
{Airapetian}, V.~S., {Glocer}, A., {Gronoff}, G., {H{\'e}brard}, E., \&
  {Danchi}, W. 2016, Nature Geoscience, 9, 452, \dodoi{10.1038/ngeo2719}

\bibitem[{{Airapetian} {et~al.}(2017){Airapetian}, {Jackman}, {Mlynczak},
  {Danchi}, \& {Hunt}}]{Airapetian17}
{Airapetian}, V.~S., {Jackman}, C.~H., {Mlynczak}, M., {Danchi}, W., \& {Hunt},
  L. 2017, Scientific Reports, 7, 14141, \dodoi{10.1038/s41598-017-14192-4}

\bibitem[{{Arumainayagam} {et~al.}(2021){Arumainayagam}, {Herbst}, {Heays},
  {Mullikin}, {Farrah}, \& {Mavros}}]{Arumainayagam21}
{Arumainayagam}, C.~R., {Herbst}, E., {Heays}, A.~N., {et~al.} 2021, arXiv
  e-prints, arXiv:2102.00094.
\newblock \doarXiv{2102.00094}

\bibitem[{{Asplund} {et~al.}(2009){Asplund}, {Grevesse}, {Sauval}, \&
  {Scott}}]{Asplund09}
{Asplund}, M., {Grevesse}, N., {Sauval}, A.~J., \& {Scott}, P. 2009, \araa, 47,
  481, \dodoi{10.1146/annurev.astro.46.060407.145222}

\bibitem[{{Barth} {et~al.}(2021){Barth}, {Helling}, {St{\"u}eken}, {Bourrier},
  {Mayne}, {Rimmer}, {Jardine}, {Vidotto}, {Wheatley}, \& {Fares}}]{Barth21}
{Barth}, P., {Helling}, C., {St{\"u}eken}, E.~E., {et~al.} 2021, \mnras, 502,
  6201, \dodoi{10.1093/mnras/staa3989}

\bibitem[{{Bourgalais} {et~al.}(2020){Bourgalais}, {Carrasco}, {Changeat},
  {Venot}, {Jovanovi{\'c}}, {Pernot}, {Tennyson}, {Chubb}, {Yurchenko}, \&
  {Tinetti}}]{Bourgalais20}
{Bourgalais}, J., {Carrasco}, N., {Changeat}, Q., {et~al.} 2020, \apj, 895, 77,
  \dodoi{10.3847/1538-4357/ab8e2d}

\bibitem[{{Cecchi-Pestellini} \& {Aiello}(1992)}]{CCP92}
{Cecchi-Pestellini}, C., \& {Aiello}, S. 1992, \mnras, 258, 125,
  \dodoi{10.1093/mnras/258.1.125}

\bibitem[{{Cecchi-Pestellini} {et~al.}(2006){Cecchi-Pestellini}, {Ciaravella},
  \& {Micela}}]{CCP06}
{Cecchi-Pestellini}, C., {Ciaravella}, A., \& {Micela}, G. 2006, \aap, 458,
  L13, \dodoi{10.1051/0004-6361:20066093}

\bibitem[{{Cecchi-Pestellini} {et~al.}(2009){Cecchi-Pestellini}, {Ciaravella},
  {Micela}, \& {Penz}}]{CCP09}
{Cecchi-Pestellini}, C., {Ciaravella}, A., {Micela}, G., \& {Penz}, T. 2009,
  \aap, 496, 863, \dodoi{10.1051/0004-6361/200809955}

\bibitem[{{Chadney} {et~al.}(2015){Chadney}, {Galand}, {Unruh}, {Koskinen}, \&
  {Sanz-Forcada}}]{Chadney15}
{Chadney}, J.~M., {Galand}, M., {Unruh}, Y.~C., {Koskinen}, T.~T., \&
  {Sanz-Forcada}, J. 2015, \icarus, 250, 357,
  \dodoi{10.1016/j.icarus.2014.12.012}

\bibitem[{{Chen} {et~al.}(2021){Chen}, {Zhan}, {Youngblood}, {Wolf},
  {Feinstein}, \& {Horton}}]{Chen21}
{Chen}, H., {Zhan}, Z., {Youngblood}, A., {et~al.} 2021, Nature Astronomy, 5,
  298, \dodoi{10.1038/s41550-020-01264-1}

\bibitem[{{Dalgarno} {et~al.}(1999){Dalgarno}, {Yan}, \& {Liu}}]{Dalgarno99}
{Dalgarno}, A., {Yan}, M., \& {Liu}, W. 1999, \apjs, 125, 237,
  \dodoi{10.1086/313267}

\bibitem[{{Erkaev} {et~al.}(2013){Erkaev}, {Lammer}, {Odert}, {Kulikov},
  {Kislyakova}, {Khodachenko}, {G{\"u}del}, {Hanslmeier}, \&
  {Biernat}}]{Erkaev13}
{Erkaev}, N.~V., {Lammer}, H., {Odert}, P., {et~al.} 2013, Astrobiology, 13,
  1011, \dodoi{10.1089/ast.2012.0957}

\bibitem[{{Fontenla} {et~al.}(2015){Fontenla}, {Stancil}, \&
  {Landi}}]{Fontenla15}
{Fontenla}, J.~M., {Stancil}, P.~C., \& {Landi}, E. 2015, \apj, 809, 157,
  \dodoi{10.1088/0004-637X/809/2/157}

\bibitem[{{Garc{\'\i}a Mu{\~n}oz}(2007)}]{Garcia07}
{Garc{\'\i}a Mu{\~n}oz}, A. 2007, \planss, 55, 1426,
  \dodoi{10.1016/j.pss.2007.03.007}

\bibitem[{{Giacobbe} {et~al.}(2021){Giacobbe}, {Brogi}, {Gandhi}, {Cubillos},
  {Bonomo}, {Sozzetti}, {Fossati}, {Guilluy}, {Carleo}, {Rainer},
  {Harutyunyan}, {Borsa}, {Pino}, {Nascimbeni}, {Benatti}, {Biazzo},
  {Bignamini}, {Chubb}, {Claudi}, {Cosentino}, {Covino}, {Damasso}, {Desidera},
  {Fiorenzano}, {Ghedina}, {Lanza}, {Leto}, {Maggio}, {Malavolta}, {Maldonado},
  {Micela}, {Molinari}, {Pagano}, {Pedani}, {Piotto}, {Poretti}, {Scandariato},
  {Yurchenko}, {Fantinel}, {Galli}, {Lodi}, {Sanna}, \& {Tozzi}}]{Giacobbe21}
{Giacobbe}, P., {Brogi}, M., {Gandhi}, S., {et~al.} 2021, \nat, 592, 205,
  \dodoi{10.1038/s41586-021-03381-x}

\bibitem[{{He} {et~al.}(2020){He}, {H{\"o}rst}, {Lewis}, {Yu}, {Moses},
  {McGuiggan}, {Marley}, {Kempton}, {Morley}, {Valenti}, \& {Vuitton}}]{He20}
{He}, C., {H{\"o}rst}, S.~M., {Lewis}, N.~K., {et~al.} 2020, Planet. Scie. J.,
  1, 51, \dodoi{10.3847/PSJ/abb1a4}

\bibitem[{{Helling} \& {Rimmer}(2019)}]{Helling19}
{Helling}, C., \& {Rimmer}, P.~B. 2019, Philosophical Transactions of the Royal
  Society of London Series A, 377, 20180398, \dodoi{10.1098/rsta.2018.0398}

\bibitem[{{Hudson} {et~al.}(2004){Hudson}, {Vallance}, \& {Harland}}]{Hudson04}
{Hudson}, J.~E., {Vallance}, C., \& {Harland}, P.~W. 2004, Journal of Physics B
  Atomic Molecular Physics, 37, 445, \dodoi{10.1088/0953-4075/37/2/012}

\bibitem[{{Huebner} \& {Mukherjee}(2015)}]{Huebner15}
{Huebner}, W.~F., \& {Mukherjee}, J. 2015, \planss, 106, 11,
  \dodoi{10.1016/j.pss.2014.11.022}

\bibitem[{{Husser} {et~al.}(2013){Husser}, {Wende-von Berg}, {Dreizler},
  {Homeier}, {Reiners}, {Barman}, \& {Hauschildt}}]{Husser13}
{Husser}, T.~O., {Wende-von Berg}, S., {Dreizler}, S., {et~al.} 2013, \aap,
  553, A6, \dodoi{10.1051/0004-6361/201219058}

\bibitem[{{Johnstone} {et~al.}(2018){Johnstone}, {G{\"u}del}, {Lammer}, \&
  {Kislyakova}}]{Johnstone18}
{Johnstone}, C.~P., {G{\"u}del}, M., {Lammer}, H., \& {Kislyakova}, K.~G. 2018,
  \aap, 617, A107, \dodoi{10.1051/0004-6361/201832776}

\bibitem[{{King} {et~al.}(2018){King}, {Wheatley}, {Salz}, {Bourrier},
  {Czesla}, {Ehrenreich}, {Kirk}, {Lecavelier des Etangs}, {Louden}, {Schmitt},
  \& {Schneider}}]{King18}
{King}, G.~W., {Wheatley}, P.~J., {Salz}, M., {et~al.} 2018, \mnras, 478, 1193,
  \dodoi{10.1093/mnras/sty1110}

\bibitem[{{Kitzmann} {et~al.}(2018){Kitzmann}, {Heng}, {Rimmer}, {Hoeijmakers},
  {Tsai}, {Malik}, {Lendl}, {Deitrick}, \& {Demory}}]{Kitzmann18}
{Kitzmann}, D., {Heng}, K., {Rimmer}, P.~B., {et~al.} 2018, \apj, 863, 183,
  \dodoi{10.3847/1538-4357/aace5a}

\bibitem[{{Koskinen} {et~al.}(2014){Koskinen}, {Lavvas}, {Harris}, \&
  {Yelle}}]{Koskinen14}
{Koskinen}, T.~T., {Lavvas}, P., {Harris}, M.~J., \& {Yelle}, R.~V. 2014,
  Philosophical Transactions of the Royal Society of London Series A, 372,
  20130089, \dodoi{10.1098/rsta.2013.0089}

\bibitem[{{Linsky} {et~al.}(2014){Linsky}, {Fontenla}, \& {France}}]{Linsky14}
{Linsky}, J.~L., {Fontenla}, J., \& {France}, K. 2014, \apj, 780, 61,
  \dodoi{10.1088/0004-637X/780/1/61}

\bibitem[{{Linsky} {et~al.}(2020){Linsky}, {Wood}, {Youngblood}, {Brown},
  {Froning}, {France}, {Buccino}, {Cranmer}, {Mauas}, {Miguel}, {Pineda},
  {Rugheimer}, {Vieytes}, {Wheatley}, \& {Wilson}}]{Linsky20}
{Linsky}, J.~L., {Wood}, B.~E., {Youngblood}, A., {et~al.} 2020, \apj, 902, 3,
  \dodoi{10.3847/1538-4357/abb36f}

\bibitem[{{Locci} {et~al.}(2018){Locci}, {Cecchi-Pestellini}, {Micela},
  {Ciaravella}, \& {Aresu}}]{Locci18}
{Locci}, D., {Cecchi-Pestellini}, C., {Micela}, G., {Ciaravella}, A., \&
  {Aresu}, G. 2018, \mnras, 473, 447, \dodoi{10.1093/mnras/stx2370}

\bibitem[{{Lorenzani} \& {Palla}(2001)}]{Lorenzani01}
{Lorenzani}, A., \& {Palla}, F. 2001, in Astronomical Society of the Pacific
  Conference Series, Vol. 243, From Darkness to Light: Origin and Evolution of
  Young Stellar Clusters, ed. T.~{Montmerle} \& P.~{Andr{\'e}}, 745.
\newblock \doarXiv{astro-ph/0011486}

\bibitem[{{Madhusudhan}(2019)}]{Madhusudhan19}
{Madhusudhan}, N. 2019, \araa, 57, 617,
  \dodoi{10.1146/annurev-astro-081817-051846}

\bibitem[{{Madhusudhan} {et~al.}(2016){Madhusudhan}, {Ag{\'u}ndez}, {Moses}, \&
  {Hu}}]{Madhusudhan16}
{Madhusudhan}, N., {Ag{\'u}ndez}, M., {Moses}, J.~I., \& {Hu}, Y. 2016, \ssr,
  205, 285, \dodoi{10.1007/s11214-016-0254-3}

\bibitem[{{Maloney} {et~al.}(1996){Maloney}, {Hollenbach}, \&
  {Tielens}}]{Maloney96}
{Maloney}, P.~R., {Hollenbach}, D.~J., \& {Tielens}, A.~G.~G.~M. 1996, \apj,
  466, 561, \dodoi{10.1086/177532}

\bibitem[{{Micela}(2002)}]{Micela02}
{Micela}, G. 2002, in Astronomical Society of the Pacific Conference Series,
  Vol. 269, The Evolving Sun and its Influence on Planetary Environments, ed.
  B.~{Montesinos}, A.~{Gimenez}, \& E.~F. {Guinan}, 107

\bibitem[{{Molaverdikhani} {et~al.}(2019){Molaverdikhani}, {Henning}, \&
  {Molli{\`e}re}}]{Molaverdikhani19}
{Molaverdikhani}, K., {Henning}, T., \& {Molli{\`e}re}, P. 2019, \apj, 883,
  194, \dodoi{10.3847/1538-4357/ab3e30}

\bibitem[{{Moses} {et~al.}(2011){Moses}, {Visscher}, {Fortney}, {Showman},
  {Lewis}, {Griffith}, {Klippenstein}, {Shabram}, {Friedson}, {Marley}, \&
  {Freedman}}]{Moses11}
{Moses}, J.~I., {Visscher}, C., {Fortney}, J.~J., {et~al.} 2011, \apj, 737, 15,
  \dodoi{10.1088/0004-637X/737/1/15}

\bibitem[{{Raymond} \& {Smith}(1977)}]{Raymond77}
{Raymond}, J.~C., \& {Smith}, B.~W. 1977, \apjs, 35, 419,
  \dodoi{10.1086/190486}

\bibitem[{{Ribas} {et~al.}(2005){Ribas}, {Guinan}, {G{\"u}del}, \&
  {Audard}}]{Ribas05}
{Ribas}, I., {Guinan}, E.~F., {G{\"u}del}, M., \& {Audard}, M. 2005, \apj, 622,
  680, \dodoi{10.1086/427977}

\bibitem[{{Sanz-Forcada} {et~al.}(2011){Sanz-Forcada}, {Micela}, {Ribas},
  {Pollock}, {Eiroa}, {Velasco}, {Solano}, \&
  {Garc{\'\i}a-{\'A}lvarez}}]{Sanz-Forcada11}
{Sanz-Forcada}, J., {Micela}, G., {Ribas}, I., {et~al.} 2011, \aap, 532, A6,
  \dodoi{10.1051/0004-6361/201116594}

\bibitem[{{Schmitt} \& {Liefke}(2004)}]{Schmitt04}
{Schmitt}, J.~H.~M.~M., \& {Liefke}, C. 2004, \aap, 417, 651,
  \dodoi{10.1051/0004-6361:20030495}

\bibitem[{{Shematovich} {et~al.}(2014){Shematovich}, {Ionov}, \&
  {Lammer}}]{Shematovich14}
{Shematovich}, V.~I., {Ionov}, D.~E., \& {Lammer}, H. 2014, \aap, 571, A94,
  \dodoi{10.1051/0004-6361/201423573}

\bibitem[{{Shulyak} {et~al.}(2020){Shulyak}, {Lara}, {Rengel}, \&
  {N{\`e}mec}}]{Shulyak20}
{Shulyak}, D., {Lara}, L.~M., {Rengel}, M., \& {N{\`e}mec}, N.~E. 2020, \aap,
  639, A48, \dodoi{10.1051/0004-6361/201937210}

\bibitem[{{Sternberg} \& {Dalgarno}(1995)}]{Sternberg95}
{Sternberg}, A., \& {Dalgarno}, A. 1995, \apjs, 99, 565, \dodoi{10.1086/192198}

\bibitem[{{Tsiaras} {et~al.}(2019){Tsiaras}, {Waldmann}, {Tinetti}, {Tennyson},
  \& {Yurchenko}}]{Tsiaras19}
{Tsiaras}, A., {Waldmann}, I.~P., {Tinetti}, G., {Tennyson}, J., \&
  {Yurchenko}, S.~N. 2019, Nature Astronomy, 3, 1086,
  \dodoi{10.1038/s41550-019-0878-9}

\bibitem[{{Venot} \& {Ag{\'u}ndez}(2015)}]{Venot15}
{Venot}, O., \& {Ag{\'u}ndez}, M. 2015, Experimental Astronomy, 40, 469,
  \dodoi{10.1007/s10686-014-9406-1}

\bibitem[{{Venot} {et~al.}(2016){Venot}, {Rocchetto}, {Carl}, {Roshni Hashim},
  \& {Decin}}]{Venot16}
{Venot}, O., {Rocchetto}, M., {Carl}, S., {Roshni Hashim}, A., \& {Decin}, L.
  2016, \apj, 830, 77, \dodoi{10.3847/0004-637X/830/2/77}

\bibitem[{{Venot} {et~al.}(2018){Venot}, {B{\'e}nilan}, {Fray}, {Gazeau},
  {Lef{\`e}vre}, {Es-sebbar}, {H{\'e}brard}, {Schwell}, {Bahrini},
  {Montmessin}, {Lef{\`e}vre}, \& {Waldmann}}]{Venot18}
{Venot}, O., {B{\'e}nilan}, Y., {Fray}, N., {et~al.} 2018, \aap, 609, A34,
  \dodoi{10.1051/0004-6361/201731295}

\bibitem[{{Verner} {et~al.}(1996){Verner}, {Ferland}, {Korista}, \&
  {Yakovlev}}]{Verner96}
{Verner}, D.~A., {Ferland}, G.~J., {Korista}, K.~T., \& {Yakovlev}, D.~G. 1996,
  \apj, 465, 487, \dodoi{10.1086/177435}

\bibitem[{{Wakelam} {et~al.}(2015){Wakelam}, {Loison}, {Herbst}, {Pavone},
  {Bergeat}, {B{\'e}roff}, {Chabot}, {Faure}, {Galli}, {Geppert}, {Gerlich},
  {Gratier}, {Harada}, {Hickson}, {Honvault}, {Klippenstein}, {Le Picard},
  {Nyman}, {Ruaud}, {Schlemmer}, {Sims}, {Talbi}, {Tennyson}, \&
  {Wester}}]{Wakelam15}
{Wakelam}, V., {Loison}, J.~C., {Herbst}, E., {et~al.} 2015, \apjs, 217, 20,
  \dodoi{10.1088/0067-0049/217/2/20}

\bibitem[{{Yan} \& {Dalgarno}(1997)}]{Yan97}
{Yan}, M., \& {Dalgarno}, A. 1997, \apj, 481, 296, \dodoi{10.1086/304034}

\end{thebibliography}
\bibliographystyle{aasjournal}

\end{document}